\documentclass[aoas,nameyear,dvips]{arximspdf}
\usepackage{algorithm,algorithmic}
\usepackage{dcolumn}
\usepackage{graphicx}

\doi{10.1214/09-AOAS286}
\volume{4}
\issue{1}
\pubyear{2010}
\firstpage{222}
\lastpage{265}
\setattribute{copyright}{owner}{\textup{In the Public Domain}}

\makeatletter

\newcolumntype{d}[1]{D{.}{.}{#1}}

\newtheorem{theorem}{Theorem}[section]
\newproclaim{assumption}[theorem]{Assumption}
\newtheorem{lemma}[theorem]{Lemma}
\newproclaim{remark}{Remark}

\def\epsilon{\varepsilon}

\newcommand{\argmax}{\operatorname{arg\max}}

\newcommand{\union}{\cup}

\newcommand{\For}[1]{\textbf{for} #1 \textbf{do}}

\newcommand{\EndFor}{\textbf{end for}}

\newcommand{\Rr}{\mathbb{R}}

\makeatother

\begin{document}
\begin{frontmatter}

\title{Sequential Monte Carlo pricing of American-style options under
stochastic volatility models\thanksref{TITL1}}
\runtitle{Sequential Monte Carlo pricing of American options}
\thankstext{TITL1}{Supported in part by equipment purchased with NSF SCREMS
Grants DMS-05-32083 and DMS-04-22400.}

\begin{aug}
\author[A]{\fnms{Bhojnarine R.} \snm{Rambharat}\thanksref{t1}\ead[label=e1]{ricky.rambharat@occ.treas.gov}\corref{}}
and
\author[B]{\fnms{Anthony E.} \snm{Brockwell}\ead[label=e2]{abrock@stat.cmu.edu}}
\thankstext{t1}{The views expressed in this paper are solely those of
the authors and do not represent the opinions
of the U.S. Department of the Treasury or the Office of the Comptroller
of the Currency. All remaining errors are the authors' responsibility.}
\runauthor{B. R. Rambharat and A. E. Brockwell}
\affiliation{U.S. Department of the Treasury
and Carnegie Mellon University}
\address[A]{Office of the Comptroller of the Currency \\
U.S. Department of the Treasury\\
250 E Street SW \\
Washington, DC 20219 \\USA\\
\printead{e1}} 
\address[B]{Department of Statistics \\
Carnegie Mellon University\\
232 Baker Hall \\
Pittsburgh, Pennsylvania 15213 \\
USA\\
\printead{e2}}
\end{aug}

\received{\smonth{1} \syear{2008}}
\revised{\smonth{6} \syear{2009}}

%
\begin{abstract}
We introduce a new method to price American-style options on
underlying investments governed by stochastic volatility (SV) models.
The method does not require the volatility process to
be observed. Instead, it
exploits the fact that the optimal decision functions in the
corresponding dynamic programming problem can be expressed
as functions of conditional distributions of volatility, given
observed data. By constructing statistics summarizing information
about these conditional distributions, one can obtain
high quality approximate solutions. Although the required conditional
distributions are in
general intractable, they can be arbitrarily precisely
approximated using sequential Monte Carlo schemes.
The drawback, as with many Monte Carlo schemes, is potentially heavy
computational demand.
We present two variants of the algorithm, one closely
related to the well-known least-squares Monte Carlo algorithm of
Longstaff and Schwartz [\textit{The Review of Financial Studies}
\textbf{14} (2001) 113--147],
and the other solving the same problem using a ``brute force''
gridding approach. We estimate an illustrative SV model using
Markov chain Monte Carlo (MCMC) methods for three equities. We also demonstrate
the use of our algorithm by estimating the posterior distribution of
the market
price of volatility risk for each of the three equities.
\end{abstract}

%

\end{frontmatter}

\section{Introduction} 
\label{sec:intro}

American-style option contracts are traded extensively over several
exchanges. These early-exercise financial derivatives are typically
written on equity stocks, foreign currency and some indices, and
include, among other examples, options on individual equities traded
on The American Stock Exchange (AMEX), options on currency traded on
the Philadelphia Stock Exchange (PHLX) and the OEX index options on
the S\&P 100 Index traded on the Chicago Board Options Exchange
(CBOE). As with any other kind of option, methods for pricing are
based on assumptions about the probabilistic model governing the
evolution of the underlying asset. Arguably, stochastic volatility
models are the most realistic models to date for underlying equities,
but existing methods for pricing American-style options have mostly
been developed using less realistic models, or else assuming that
volatility is observable. In this paper we develop a new method for
pricing American-style options when the underlying process is
governed by a stochastic volatility model, and the volatility is not
directly observable. The method yields near-optimal solutions under
the model assumptions, and can also formally take into account the market
price of volatility risk (or volatility risk premium).

It follows from the fundamental theorem of arbitrage that an option
price can be determined by computing the discounted expectation of the
payoff of the option under a risk-neutral measure, assuming that the
exercise decision is made so as to maximize the payoff. While this is
simple to compute for European-style options, as illustrated in the
celebrated papers of \citet{bs73} and \citet{mer73}, the pricing
problem is
enormously more difficult for American-style options, due to the
possibility of early exercise. For American-style options, the price
is in fact the supremum over a large range of possible stopping times
of the discounted expected payoff under a risk-neutral measure. A
range of methods has been developed to find this price, or
equivalently, to solve a corresponding stochastic dynamic programming
problem. \citet{glasserman} provides a thorough review of
American-style option pricing with a strong emphasis on Monte Carlo
simulation-based procedures. Due to the difficulty of the problem,
certain assumptions are usually made. For instance, a number of
effective algorithms [including those developed in \citet{bs77}, \citet{bg97}, \citet{cjm92}, \citet{gj84},
\citet{LongSch},
\citet{rog02},  and
\citet{sullivan00}] are based on the assumption that the underlying asset
price is governed by a univariate diffusion process with a constant
and/or directly observable volatility process.

As recognized by \citet{bs73} and others, the assumption of constant
volatility is typically inconsistent with observed data. The volatility
``smile'' (or ``smirk'') is one example where empirical data show
evidence against constant volatility models. The smile (smirk) effect
arises when option contracts with different strike prices, all other
contract features being equivalent, result in different implied
volatilities (i.e., the volatility required to calibrate to market
observed option prices).
A variety of more realistic models has subsequently been developed for asset
prices, with stochastic volatility models arguably representing the
best models to date. This has led researchers to develop pricing
methods for European-style options
when the underlying asset price is governed by stochastic volatility
models [e.g., \citet{fps00}, \citet{hes93}, \citet{hw87} and \citet{ss91}]. However, work on pricing of American-style options under
stochastic volatility models is far less developed. A number of
authors [including \citet{cp99}, \citet{fintom97}, \citet{fps00}, \citet{guanguo}, \citet{tw03} and \citet{ZhangLim06}]
have made valuable inroads in addressing this problem, but most assume that
volatility is observable. \citet{fps00} provide an
approximation scheme based on the assumption of fast mean-reversion
in the volatility process, and they use a clever asymptotic expansion
method to correct the
constant volatility option price to account for stochastic volatility.
The correction involves parameters estimated from the implied
volatility surface and they derive a pricing equation that does not
depend directly on the volatility process. \citet{tw03} use an
analytic-based approach whereby they compute the optimal stopping
boundary using Chebyshev polynomials to value American-style options
in a stochastic volatility framework. They derive an integral
representation of the option price that depends on both the share price
and level of volatility.

Additionally, the approach in \citet{cp99}
uses a multi-grid technique where both the asset price and volatility
are state variables in a two-dimensional parabolic partial
differential equation (PDE). Pricing options in a stochastic
volatility framework using PDE methods are feasible once we assume that
volatility itself is an observed state variable. Typically, there is
a grid in one dimension for the share price and another dimension for
the volatility. The final option price is a function of both the
share price and volatility. \citet{guanguo} derive a lattice-based
solution to pricing American options with stochastic volatility.
Following the work in \citet{fintom97}, they construct a
lattice that depends directly on the asset price and volatility and
they illustrate an empirical study where they back out the parameters
from the stochastic volatility model using data on American-style
S\&P 500 futures options. The valuation algorithm that they develop,
however, involves an explicit dependence on both the share price and
volatility state variables. Additionally, \citet{ZhangLim06} propose a
valuation approach that is based on a decomposition of American option
prices, however, volatility is a variable in the pricing result.

The approach we develop, in contrast with the aforementioned methods,
combines the optimal decision-making problem with the volatility
estimation problem. We assume that the asset price follows a stochastic
volatility model, that observations are made at discrete points in
time $t=0,1,2,\ldots,$ and that exercise decisions are made
immediately after each observation. It could be argued that
volatility should be considered observable since one could simply
compute ``Black and Scholes type'' implied volatilities from observed
option prices.
However, implied volatilities are based on the assumption of a simple geometric
Brownian motion model (or some other simplified diffusion process) and,
thus, their use would defeat the purpose of developing pricing methods
with more realistic models.
The implied volatility calculation is not as straightforward in a
stochastic volatility (multivariate) modeling framework. \citet{rentou96} approximate a Black and Scholes type implied volatility
quantity in a \citet{hw87} setting. The analysis in \citet{rentou96}
computes filtered volatilities using an iterative approach and
illustrates applications to hedging problems.
On the other hand, we aim to compute the posterior distribution of
volatility conditional on observed data. Our pricing scheme is based on
two key observations. First, we use a
sequential Monte Carlo (also referred to as ``particle filtering'')
scheme to perform inference on the unobserved volatility process at
any given point in time. Second, conditional distributions of the
unobserved volatility at a given point in time, given current and past
observations of the price process, which are necessary for finding an
exact solution to the dynamic programming problem, can be
well approximated by a summary vector or by a low-dimensional
parametric family of distributions.

Inference on the latent volatility process for the purpose of option
pricing is an application of the more general methodology that
addresses partially observed time series models in a dynamic
programming/optimal control setting
[see, e.g., \citeauthor{Bertsekas00a} (\citeyear{Bertsekas00a,Bertsekas00b}) and the references
therein]. Among earlier work, \citet{sorenstubb68} provide a method
based on Edgeworth expansions to estimate the posterior density of the
latent process in a nonlinear, non-Gaussian state-space modeling
framework. Our objective in this paper is to illustrate an algorithm
that allows an agent (a holder of an American option) to optimally
decide the exercise time assuming that both the share price and
volatility are stochastic variables. In our partially observed setting,
only the share price is observable; volatility is a latent process. The
main challenge for the case of American-style options is determining
the continuation value so that an optimal exercise/hold decision can be
made at each time point.

Several researchers have applied Monte Carlo methods
to solve the American option pricing problem. Most notable among
these include \citet{LongSch}, \citet{tvr01}, \citet{bg97} and
\citet{car96}. An excellent summary of the
work done on Monte Carlo methods and American option pricing is
presented in Chapter 8 of \citet{glasserman}. The
least-squares Monte Carlo (LSM) algorithm of \citet{LongSch} has achieved
much popularity because of its intuitive regression-based approach to
American option pricing.
The LSM algorithm is very efficient to price American-style options
since as long as
one could simulate observations from the pricing model, then a
regression-based procedure could be employed along with the dynamic
programming algorithm to price the options. Pricing
American-style options in a stochastic volatility framework is
straightforward using the methodology of \citet{LongSch} as long as
draws from the share price and volatility processes can be obtained.
That is, at each time point, $n$, the decision of whether or not to
exercise an American option will be a function of the time $n$ share
price, $S_n$ (and possibly some part of its recent history $S_{n-1}$,
$S_{n-2},\ldots$), and the time $n$ volatility, $\sigma_n$.
Our pricing framework, however, needs to accommodate a
\textit{latent} volatility process.

We propose to combine the \citet{LongSch} idea with a sequential Monte
Carlo step whereby at each time point $n$, we estimate the conditional
(posterior) distribution of the latent volatility given the observed
share price data up to that time. We propose a Monte Carlo based
approach that uses a summary vector to capture the key features of
this conditional distribution. As an alternative, we also explore a
grid-based approach, studied in \citet{RambharatPhD}, whereby we propose
a parametric approximation to the conditional distribution that is
characterized by the summary
vector. Therefore, our exercise decision at a given time point is
based on the observed share price and the computed summary vector
components at that time.
Although our pricing approach is computationally intensive, as it
combines nonlinear filtering with the early-exercise feature of
American-style option valuation, it provides a way to solve the
associated optimal stopping problem in the presence of a latent
stochastic process.
We compare our approach to a basic LSM
method whereby at a given time point, we use a few past observations
to make the exercise decision in order to price American-style options
in a stochastic volatility framework.
We also compare our method to the LSM approach using the current share
price and an estimate of the current realized volatility as a proxy for
the true volatility. In order to assess precision, we compare LSM with
past share prices, LSM with realized volatility, and our proposed
valuation technique to the LSM-based American option price assuming
that share price and volatility can be observed (the full information
state). The method closest to the full information state benchmark
would be deemed the most accurate.

The present analysis addresses the problem of pricing an American-style option,
once a good model has already been found. It is worth noting, however,
that since neither
the sequential Monte Carlo scheme nor the gridding approach used in
our pricing technique are tied to a particular model, the method is
generalizable in a straightforward manner to handle a fairly wide
range of stochastic volatility models. Thus, it could be used to
perform option pricing under a range of variants of stochastic
volatility models, such as those discussed in~\citet{Chernovetal}.
Although the focus of our paper is not model estimation/selection
methodology, we do implement a thorough statistical exercise using
share price history to estimate model parameters from a stochastic
volatility model. [See
\citet{cg00}, \citet{Eraker}, \citet{Gallantetal},
\citet{GhyselsHarveyRenault},
\citet{Jacquieretal},
\citet{KimShephardChib}
and   \citet{pan02} for examples of work that mainly address model
estimation/selection issues].
Our approach will be to employ a sequential Monte Carlo based procedure
to estimate the log-likelihood of our model [see, e.g., \citet{KitagawaSato} and the references therein] and then use a Markov chain
Monte Carlo (MCMC) sampler to obtain posterior distributions of all
model parameters. Conditional on a posterior summary measure of the
model parameters (such as the mean or median), we then estimate the
(approximate) posterior distribution of the market price of volatility
risk. Our analysis is illustrated in the context of three equities
(Dell Inc., The Walt Disney Company and Xerox Corporation).

The paper is organized as follows. In Section~\ref{sec:svmodel} we
formally state
the class of stochastic volatility models with which we work. In
Section~\ref{sec:dynprog} we review the dynamic programming approach
for pricing
American-style options and demonstrate how it can be transformed into
an equivalent form, and introduce (i) a sequential Monte Carlo scheme
that yields certain conditional distributions, and (ii) a gridding
algorithm that makes use of the sequential Monte Carlo scheme to
compute option prices. Section~\ref{sec:pricealg} describes the pricing
algorithms and presents some illustrative numerical experiments.
In Section~\ref{sec:statmethod} we describe (i) the MCMC estimation
procedure for the stochastic volatility model parameters, and (ii) the
inferential analysis of the market price of volatility risk.
Section~\ref{sec:dataexercise} contains posterior results of our
empirical analysis with observed market data. Section~\ref{sec:discuss}
provides concluding remarks.
Finally, we present additional technical details in the
\hyperref[sec:appendix]{Appendix}.
We also state references to our computing
code and data sets in supplementary.


\section{The stochastic volatility model} 
\label{sec:svmodel}
\setcounter{footnote}{2}
Let $(\Omega, \mathcal{F}, P)$ be a probability space,
and let $\{S(t),~t \ge0\}$ be a stochastic process defined
on $(\Omega, \mathcal{F}, P)$, describing the evolution of
our asset price over time. Time $t=0$ will be referred to
as the ``current'' time, and we will be interested in an option
with expiry time $(T \Delta)>0$, with $T$ being some positive integer,
and $\Delta$ some positive real-valued constant.
Assume that we observe the process only
at the discrete time points $t=0,\Delta,2\Delta,\ldots,T\Delta$,
and that exercise
decisions are made immediately
after each observation.
(We would typically take the time unit here
to be one year, and $\Delta$ to be $1/252$, representing one trading day.
However, both $\Delta$ and the time units can be chosen arbitrarily
subject to the constraints mentioned above.)

Assume that, under a risk-neutral measure, the asset price $S(t)$
evolves according to the It{\^{o}} stochastic differential equations
(SDEs)\footnote{Under the statistical (or real-world) measure, the
asset price evolves on another probability space. Under the real-world
measure, the drift term $r$ in equation (\ref{eq:rnsvmodelall1}) is
replaced by the physical drift and the term $\frac{\lambda\gamma
}{\alpha}$ does not appear in the drift of equation (\ref
{eq:rnsvmodelall3}). The change of measure between real-world and
risk-neutral is formalized through a Radon--Nikodym derivative.}
%
\begin{eqnarray}
\label{eq:rnsvmodelall1}
dS(t) &=& r S(t) \,dt + \sigma(Y(t)) S(t) \bigl[\sqrt{1-\rho^2} \,dW_1(t) +
\rho\,dW_2(t) \bigr],
\\
\label{eq:rnsvmodelall2}
\sigma(Y(t)) &= &\exp (Y(t) ),
\\
\label{eq:rnsvmodelall3}
dY(t) &=& \biggl[\alpha\biggl(\beta- \frac{\lambda\gamma}{\alpha}- Y(t) \biggr) \biggr]\, dt +
\gamma \,dW_2(t),
\end{eqnarray}
where $r$ represents the risk-free interest rate (measured in
appropriate time units), $\sigma(Y(t))$
is referred to as the ``volatility,'' $\rho$ measures the co-dependence
between the share price and volatility processes, $\alpha$ (volatility
mean reversion rate), $\beta$ (volatility mean reversion level), and
$\gamma$ (volatility of volatility)
are constants with $\alpha>0$, $\gamma>0$, $\{W_1(t)\}$ and $\{
W_2(t)\}
$ are
assumed to be two independent standard Brownian motions, and $\lambda$
is a constant referred to as the ``market price of volatility risk'' or
``volatility risk premium'' [\citet{bk03a}, \citet{mw01} and \citet{mr98}]. If we set
\[
dW_1^*(t) = \bigl[\sqrt{1-\rho^2} \,dW_1(t) + \rho \,dW_2(t) \bigr],
\]
we can more clearly see that $\rho$ is the correlation between the
Brownian motions $dW_1^*(t)$ and $dW_2(t)$. The parameter $\rho$
quantifies the so-called ``leverage effect'' between share prices and
their volatility.

Observe that $\lambda$ is not uniquely determined in the above system
of SDEs. Since we are working in a stochastic volatility modeling
framework, markets are said to be ``incomplete'' because volatility is
not a traded asset and cannot be perfectly hedged. This is to be
contrasted with the constant volatility \citet{bs73} framework where a
unique pricing measure exists and all risks can be perfectly hedged away.
There is a range of possible risk-neutral measures when pricing under
stochastic volatility models, each one having a different value of
$\lambda$. In fact, there are also
risk-neutral measures under which $\lambda$ varies over time.
However, for the sake of simplicity, we will assume that $\lambda$
is a constant. Later in this paper,
we illustrate how to estimate $\lambda$.

Equations (\ref{eq:rnsvmodelall1})--(\ref{eq:rnsvmodelall3}) represent a stochastic volatility model that accommodates
mean-reversion in volatility and we will use it to
illustrate our American option valuation methodology. It is an example
of a nonlinear,
non-Gaussian state-space model. \citet{sco87} represents one of the
earlier analyses of this stochastic volatility model. This same type of
model has also been studied
in a Bayesian context by \citet{Jacquieretal} and, more recently, in
\citet{jpr04} and \citet{junyu05}. It should be noted that our
methodology is not constrained to a specific stochastic volatility
model. The core elements of our approach would apply over a broad
spectrum of stochastic volatility models such as, for instance, the
\citet{hw87} and \citet{hes93} stochastic volatility models.

Since our observations occur
at discrete time-points $0,\Delta,2\Delta,\ldots,$
we will make extensive
use of the discrete-time approximation
to the solution of the risk-neutral stochastic differential
equations~(\ref{eq:rnsvmodelall1})--(\ref{eq:rnsvmodelall3}) given by
%
\begin{eqnarray}
\label{eq:dtapprox1}
\quad S_{t+1} &=& S_{t} \cdot\exp \biggl\{ \biggl(r -
\frac{\sigma_{t+1}^2}{2} \biggr) \Delta+ \sigma_{t+1} \sqrt{\Delta} \bigl[
\sqrt{1-\rho^2} Z_{1,t+1} + \rho Z_{2,t+1} \bigr] \biggr\}, \\
\label{eq:dtapproxvol}
\sigma_{t+1} &=& \exp(Y_{t+1}), \\
\label{eq:dtapprox2}
Y_{t+1} &=& \beta^* + e^{- \alpha\Delta} (Y_{t} - \beta^*) + \gamma
\sqrt
{ \biggl(\frac{1 - e^{-2 \alpha\Delta}}{2 \alpha} \biggr)} Z_{2,t+1},
\end{eqnarray}
where $\{Z_{i,t}\}$, $i=1, 2$, is an
independent and identically distributed (i.i.d.) sequence of random
variables with standard normal [$N(0,1)$] distributions,
\[
\beta^* = \beta- \frac{\lambda\gamma}{\alpha},
\]
and all other parameters are as previously defined. Thus, $S_t$ and
$Y_t$ represent approximations, respectively, to $S(t\Delta)$ and
$Y(t\Delta)$. [The expression for $Y_t$ is obtained directly from the
exact solution to~(\ref{eq:rnsvmodelall3}), while the expression for
$S_t$ is the solution to~(\ref{eq:rnsvmodelall1}) that one would obtain
by regarding $\sigma_t$ to be constant on successive intervals of
length $\Delta$. The approximation for $S_t$ becomes more accurate as
$\Delta$ becomes smaller.]

It will sometimes be
convenient to express~(\ref{eq:dtapprox1}) in terms of the log-returns
$R_{t+1} = \log(S_{t+1}/S_{t})$, as
%
\begin{equation}
\label{eq:dtapprox3}
R_{t+1} = \biggl(r - \frac{\sigma_{t+1}^2}{2} \biggr) \Delta+ \sigma_{t+1}
\sqrt
{\Delta} \bigl(\sqrt{1-\rho^2} Z_{1, t+1} + \rho Z_{2,t+1} \bigr) .
\end{equation}
To complete the specification of the model, we can assign
$Y_0$ a normal distribution,
%
\begin{equation}
\label{eq:initydn}
Y_0 \sim N \biggl(\beta^*, \frac{\gamma^2}{2 \alpha} \biggr).
\end{equation}
This is simply the stationary (limiting) distribution of the
first-order autoregressive process $\{Y_t\}$.
However, for practical purposes, it will usually be preferable to
replace this distribution by the conditional distribution of $Y_0$,
given some historical observed price data $S_{-1},S_{-2},\ldots.$
Another reasonable starting value for $Y_0$ is a historical volatility
based measure (i.e., the log of the standard deviation of a few past
observations).
Additionally, there are examples of stochastic volatility models where
exact simulation is not feasible. In such cases, we must resort to
numerical approximation schemes such as the Euler--Maruyama method (or
any other related approach). \citet{KloedenPlaten00} illustrate several
numerical approximation schemes that could be applied to simulate from
a stochastic volatility model where no exact simulation methodology exists.


\section{Dynamic programming and option pricing} 
\label{sec:dynprog}

The arbitrage-free price of an American-style option is
%
\begin{equation}
\label{eq:amop}
\sup_{\tau\in\mathcal{T}} E_{\mathrm{RN}}[\exp(-r\tau) g(S_\tau)],
\end{equation}
where $\tau$ is a random stopping time at which an exercise
decision is made, $\mathcal{T}$ is the set of all possible
stopping times with respect to the filtration $\{\mathcal
{F}_t,~t=0,1,\ldots\}$
defined by
\[
\mathcal{F}_t = \sigma(S_0,\ldots,S_t),
\]
$E_{\mathrm{RN}}(\cdot)$ represents the expectation, under a risk-neutral
probability measure, of its argument, and $g(s)$ denotes the payoff
from exercise of the option when the underlying asset price is equal
to $s$. For example, a call option with strike price $K$ has payoff
function $g(s) = \max(s-K,0)$, and a put option with strike price $K$
has payoff function $g(s) = \max(K-s,0)$. [The analysis in this paper
is in the context of American put options since these options typically
serve as canonical examples of early-exercise derivatives; see \citeauthor{ks91} (\citeyear{ks91,ks98}) and \citet{myn92}  for key mathematical results
concerning American put options.]
Since $\tau$ is a stopping
time, the event $\{\tau\le t\}$ must be $\mathcal{F}_t$-measurable, or
equivalently, the decision to exercise or hold at a given time must be
made only based on observations of the previous and current values of
the underlying price process. To allow for the possibility that the
option is never exercised, we adopt the convention that $\tau=\infty$
if the option is not exercised at or before expiry, along with the
convention that $\exp(-r\infty) g(S_\infty)=0.$
In order to price the American option, we need to find
the stopping time $\tau$ at which the supremum
in~(\ref{eq:amop}) is achieved.
While it is not immediately obvious how one might
search through the space of all possible stopping times,
this problem is equivalent to a stochastic control problem,
which can be solved (in theory) using the dynamic programming
algorithm, which was originally developed by \citet{bellman}.
\citet{SheldonRoss1983} also describes some key principles of stochastic
dynamic programming and a thorough treatment is presented, in the
context of financial analysis, by \citet{glasserman}.

Our objective is to find the optimal stopping rule (equivalently, the
optimal exercise time) while taking into account the latent stochastic
volatility process. The key difficulty arises because the price itself
is not Markovian. We would like to get around this by using the fact
that the bivariate process, composed of both price and volatility, is
Markovian, but unfortunately we can only observe one component of that
process. It is known that one can still find the optimal stopping rule
if one can determine (exactly) the conditional distribution of the
unobservable component of the Markov process, given current and past
observable components [see, e.g., \citet{DeGroot1970} and \citet{SheldonRoss1983}]. In such a case, we can use algorithms like the ones
described in \citet{BrockwellKadane} to find the optimal decision rules.
These algorithms effectively integrate the utility function at each
point in time over the distribution of unknown quantities given
observed quantities. Unfortunately, in the context of this paper, we
cannot obtain the required conditional distributions exactly, but we
can find close approximations to them. We will therefore approach our
pricing problem by using numerical algorithms in the style of \citet{BrockwellKadane}, in conjunction with close approximations to the
required distributions. In doing so, we make the assumption (stated
later in this paper) that our distributional approximations are close
to the required conditional distributions.

\subsection{General method}
\label{subsec:0}

The dynamic programming algorithm
constructs the exact
optimal decision functions recursively, working its way from the
terminal decision point (at time $T\Delta$) back to the
first possible decision point (at time $0$).
In addition, the procedure yields the expectation
in the expression~(\ref{eq:amop}), which is our desired option price.
The algorithm works as follows.

Let $d_t \in\{E,H\} $ denote the decision made immediately after
observation of $S_t$, either to exercise ($E$) or hold ($H$) the option.
While either decision could be made, only one is optimal,
given available information up to time~$t$.
(In the event that both are optimal, we will assume that the agent will
exercise the option.)
We denote the optimal decision, as a function of the available
observations, by
\[
d_t^*(s_0,\ldots,s_t) \in\{E,H\}.
\]
Here and in the remainder of the paper, we adopt the usual convention
of using $S_j$ (upper case)
to denote the random variable representing the equity
price at time~$j$, and $s_j$ (lower case) to denote a particular
possible realization of the random variable.
Next, let
%
\begin{equation}
\label{eq:uTdefn}
u_T(s_0,\ldots,s_T,d_T) =
\cases{
g(s_T), &\quad $d_T = E$, \cr
0, &\quad $d_T = H$,
}
\end{equation}
and for $t=0,1,\ldots,T-1$,
let $u_t(s_0,\ldots,s_t,d_t)$ denote the discounted expected payoff
of the option at time $(t \Delta)$, assuming that decision $d_t$ is made,
and \emph{also assuming} that optimal decisions are made at times
$(t+1)\Delta,(t+2)\Delta,\ldots,T \Delta.$
It is obvious that, at the expiration time,
\[
d_T^*(s_0,\ldots,s_T) = \argmax\limits_{d_T \in\{E,H\}}
u_T(s_0,\ldots,s_T,d_T).
\]
The optimal decision functions $d^*_{T-1},\ldots,d^*_0$ can
then be obtained by defining
%
\begin{equation}
\label{eq:utstardefn}
u_t^*(s_0,\ldots,s_t) =
u_t(s_0,\ldots,s_t,d_t^*(s_0,\ldots,s_t)),\qquad t=0,\ldots,T,
\end{equation}
and using the recursions
\begin{eqnarray}
\label{eq:dprecursion}
&&u_t(s_0,\ldots,s_t,d_t) \nonumber\\[-8pt]\\[-8pt]
&&\qquad=
\cases{
g(s_t), &\quad $d_t = E$, \cr
E_{\mathrm{RN}}\bigl({u_{t+1}^*(S_0,\ldots,S_{t+1}) | S_0=s_0,\ldots,S_t=s_t}\bigr), &\quad $d_t
= H$,
}\nonumber
\end{eqnarray}
\begin{equation}
\label{eq:dtstardefn}
d_t^*(s_0,\ldots,s_t) =
\argmax\limits_{d_t \in\{E,H\}} u_t(s_0,\ldots,s_t,d_t).
\end{equation}
These recursions are used sequentially, for
$t=T-1,T-2,\ldots,0,$
and yield the (exact) optimal decision
functions $d^*_t,~t=0,\ldots,T .$ (Each $d_t$ is optimal in the
space of all possible functions of historical data $s_0,\ldots,s_t.$)
The corresponding stopping time $\tau$ is simply
\[
\tau= \min\bigl( \bigl\{ t \in\{0,\ldots,T\} | d_t=E \bigr\} \union\{\infty\} \bigr).
\]
Furthermore, the procedure also gives the risk-neutral option price, since
\[
u^*_0(s_0) = \sup_{\tau\in\mathcal{T}} E_{\mathrm{RN}}[\exp(-r\tau)
g(S_\tau)].
\]
In practice, it is generally not possible to compute the optimal
decision functions, since each $d^*_t$ needs to be computed and stored
for all (infinitely many) possible combinations of values of its
arguments $s_0,\ldots,s_t.$ However, in what follows we will develop
an approach which gives high-quality approximations to the exact
solution. The approach relies on exploiting some key features of the
American-style option pricing problem.

\subsection{Equivalent formulation of the dynamic programming problem}
\label{subsec:1}

First, it follows from the Markov property of
the bivariate process $\{(S_t,Y_t)\}$ that we
can transform the arguments of the decision functions
so that they do not increase in number as $t$ increases.
Let us define
%
\begin{equation}
\label{eq:pitdefn}
\pi_t(y_t) \,dt = P(Y_t \in dy_t | S_0=s_0,\ldots,S_t=s_t),\qquad
t=0,\ldots,T,
\end{equation}
where $s_j$ denotes the observed value of $S_j$,
so that $\pi_t(\cdot)$ denotes the conditional
density of the distribution of $Y_t$
(with respect to the Lebesgue measure), given
historical information $S_0=s_0,\ldots,S_t=s_t$.
Then we have the following result.

\begin{lemma}\label{lem:1} 
For each $t=0,\ldots,T$, $u_t(s_0,\ldots,s_t,d_t)$
can be expressed as a functional of only
$s_t$, $\pi_t$ and $d_t$, that is,
\[
u_t(s_0,\ldots,s_t,d_t) =
\tilde{u}_t(s_t,\pi_t,d_t).
\]
Consequently, for each $t=0,\ldots,T$,
$d_t^*(s_0,\ldots,s_t)$ and $u^*_t(s_0,\ldots,s_t)$
can also be expressed, respectively, as functionals
\begin{eqnarray*}
d_t^*(s_0,\ldots,s_t) & = & \tilde{d}_t^*(s_t,\pi_t), \\
u^*_t(s_0,\ldots,s_t) & = & \tilde{u}_t^*(s_t,\pi_t).
\end{eqnarray*}
\end{lemma}

This is a special case of the well-known result [see, e.g., \citeauthor{Bertsekas00a} (\citeyear{Bertsekas00a,Bertsekas00b})] on the sufficiency of
filtering distributions in optimal control problems. A proof is given
in the \hyperref[sec:appendix]{Appendix}. It is important to note here that the argument
$\pi_t$ to the functions $\tilde{u}_t$, $\tilde{d}_t^*$ and
$\tilde{u}_t^*$ is a function itself.

Lemma~\ref{lem:1} states that each optimal decision function can be
expressed as a functional depending only on the current price $s_t$
and the conditional distribution of $Y_t$ (the latent process that
drives volatility)
given observations of prices $s_0,\ldots,s_t .$
In other words, we can write the exact equivalent form of
the dynamic programming recursions,
%
\begin{eqnarray}
\label{eq:dpequiv01}
\quad\tilde{u}_T(s_T,\pi_T,d_T) &=& \cases{
g(s_T), &\quad $d_T = E$, \cr
0, &\quad $d_T = H$,
}
\\
\label{eq:dpequiv02}
\tilde{u}_t(s_t, \pi_t, d_t) &=& \cases{
g(s_t), &\quad $d_t = E$, \cr
E_{\mathrm{RN}}[ \tilde{u}_{t+1}^*(S_{t+1},\pi_{t+1}) | S_t=s_t, \pi_t],
&\quad
$d_t = H$,
}
\end{eqnarray}
where
$
\tilde{d}_t^*(s_t,\pi_t) = \argmax_{d_t \in\{E,H\}} \tilde
{u}_t(s_t,\pi_t,d_t),
$
and
$
\tilde{u}_t^*(s_t,\pi_t) = \tilde{u}_t(s_t,\pi_t,\break\tilde
{d}_t^*(s_t,\pi_t)).
$

\subsection{Summary vectors and sequential Monte Carlo}
\label{subsec:2}
In order to implement the ideal dynamic programming algorithm as laid out
in Section~\ref{subsec:1}, we would need (among other things)
to be able to determine the filtering distributions
$\pi_t(\cdot),~t=0,1,\ldots,T.$
Unfortunately, since these distributions are themselves infinite-dimensional
quantities in our modeling framework, we cannot work directly with
them. We can, however,
recast the dynamic programming problem in terms of $l$-dimensional
summary vectors
%
\begin{equation}
\label{eq:qt}
Q_t = \left[
\matrix{
f_1(\pi_t) \cr
\vdots\cr
f_l(\pi_t)
}
\right],
\end{equation}
where $f_1(\cdot),\ldots,f_l(\cdot)$ are some functionals.
The algorithms we introduce can be used with any choice of these functionals,
but it is important that they capture key features of the distribution.
As typical choices, one can use the moments of the distribution.
In the examples in this paper, we take $l=2$,
$f_1(\pi_t) = \mathbf{E} [ \pi_t ] = \int x \pi_t(x) \,dx$ and
$f_2(\pi_t) = \operatorname{std.dev.}(\pi_t).$
Adding components to this vector typically provides more
comprehensive summaries of the required distributions,
but incurs a computational cost in the algorithms.

%
\begin{algorithm}[b]
\caption{Sequential Monte Carlo estimation of $\pi_0,\ldots,\pi_T$}\label{alg:smc}
\begin{algorithmic}[0]
\STATE\textit{Initialization ($t=0$).} Choose a number of ``particles'' $m>0$.
Draw a sample $\{\tilde{y}_0^{(1)},\ldots,\tilde{y}_0^{(m)}\}$ from
the distribution of $Y_0$ [see equation~(\ref{eq:initydn})].\\
\vskip2mm
\For{$t=1,\ldots,T$}
\begin{itemize}
\item \textit{Step 1: Forward simulation.}
For $i=1,2,\ldots,m$, draw $\tilde{y}_{t}^{(i)}$ from the distribution
$p(y_{t} | Y_{t-1}={y}_{t-1}^{(i)})$. [This distribution is
Gaussian in our example, specified by~(\ref{eq:dtapprox2}).]
\item
\textit{Step 2: Weighting.}
Compute the weights
%
\begin{equation}
\label{eq:basicweights}
w_{t}^{(i)} = p\bigl(r_{t} | Y_{t} = \tilde{y}_{t}^{(i)}\bigr),
\end{equation}
where the term on the right denotes the conditional density of the
log-return $R_{t}$, given
$Y_{t} = \tilde{y}_{t}^{(i)}$, evaluated at the observed value $r_{t}.$
[These weights are readily obtained from~(\ref{eq:dtapprox3}).]
\item
\textit{Step 3: Resampling.}
Draw a new sample
$\{{y}_{t}^{(1)},\ldots,{y}_{t}^{(m)}\}$ by
sampling with replacement from
$\{\tilde{y}_{t}^{(1)},\ldots,\tilde{y}_{t}^{(m)}\}$,
with probabilities proportional to $w_{t}^{(1)},\ldots,w_{t}^{(m)}$.
\end{itemize}
\EndFor
\end{algorithmic}
\end{algorithm}

Ultimately, we need to be able to
compute $\pi_t(\cdot)$ and thus $Q_t$. One way to do this
is with the use of sequential Monte Carlo
simulation [also known as
``particle filtering,'' see, e.g., \citet{dfg01}, for discussion,
analysis and
examples of these algorithms].
The approach yields samples
from (arbitrarily close approximations to) the distributions $\pi_t$,
and thus allows us to evaluate components of $Q_t$.
A full treatment of sequential Monte Carlo methods is
beyond the scope of this paper, but the most basic form of the
method, in this context, appears in Algorithm~\ref{alg:smc}.

This algorithm yields $T+1$ collections of particles,
$\{y_t^{(1)},\ldots,y_t^{(m)}\},~t=0,1,\ldots,T,$ with the
property that for each $t$,
$\{y_t^{(1)},\ldots,y_t^{(m)}\}$ can be regarded
as an approximate sample of size $m$ from the distribution $\pi_t$.
The algorithm has the convenient property that as
$m$ increases, the empirical distributions of
the particle collections converge to the desired
distributions $\pi_t$. Typically $m$ is chosen to be as large
as possible subject to computational constraints; for the algorithms
in this paper, we choose $m$ to be around $500$. There are several ways
to improve the sampling efficiency of Algorithm~\ref{alg:smc}, namely,
ensuring that the weights in equation~(\ref{eq:basicweights}) do not
lead to degeneration of the particles.
[For more details and various refinements of this algorithm,
refer to, among others, \citet{KitagawaSato}, \citet{LiuWest} and \citet{PittShephard}.]

\subsection{Approximate dynamic programming solution}
\label{subsec:3}

In order to motivate our proposed American option pricing methodology,
we state an assumption that describes how $Q_t$ relates to past price
history captured in $\pi_t$.

\begin{assumption}\label{assumption:qt}
The summary vector $Q_t$ is ``close to sufficient,'' that is, it
captures enough information from the past share price history\break $(S_t,
S_{t-1}, \ldots)$, so that $p(S_{t+1}|Q_t)$ is close to $p(S_{t+1}|S_t,
S_{t-1}, S_{t-2}, \ldots)$.
\end{assumption}

(If the summary vector was sufficient, then the dynamic programming
algorithm would yield exact optimal decision rules. Of course, even in
this ideal case, the numerical implementation of a dynamic programming
algorithm introduces some small errors.)

It is beyond the scope of this paper to quantify the ``closeness''
between $p(S_{t+1}|Q_t)$ and
$p(S_{t+1}|S_t, S_{t-1}, S_{t-2}, \ldots)$. One could, however,
theoretically do so using standard distribution distance measures and
perform an analysis of the propagation of the error through the
algorithms discussed in this paper.

Combining the equivalent form of the dynamic programming
recursions~(\ref{eq:dpequiv01}), (\ref{eq:dpequiv02}) along with
Assumption~\ref{assumption:qt},
we can approximate the
dynamic programming recursions by
%
\begin{equation}\label{eq:terminal}
\hat{u}_T(s_T, Q_T, d_T) = \cases{
g(s_T), &\quad $d_T = E$, \cr
0, &\quad $d_T = H$,
}
\end{equation}
\begin{eqnarray}
\label{eq:mainrecursion}
&&\hat{u}_t(s_t, Q_t, d_t)\nonumber\\[-8pt]\\[-8pt]
&&\qquad= \cases{
g(s_t), &\quad $d_t = E$, \cr
E_{\mathrm{RN}}[ \hat{u}_{t+1}^*(S_{t+1},Q_{t+1})
| S_t=s_t, Q_t=q_t], &\quad $d_t = H$, }\nonumber
\end{eqnarray}
with
%
\begin{equation}
\label{eq:dstar}
\hat{d}_t^*(s_t, Q_t)
= \argmax\limits_{d_t \in\{E,H\}} \hat{u}_t(s_t,Q_t,d_t)
\end{equation}
and
%
\begin{equation}
\label{eq:ustar}
\hat{u}_t^*(s_t, Q_t) =
\hat{u}_t(s_t, Q_t,\hat{d}_t^*(s_t, Q_t)),
\end{equation}
where $Q_t$ is the vector summarizing $\pi_t(\cdot)$ [cf.
equation~(\ref
{eq:qt})].
The recursion~(\ref{eq:mainrecursion}) is convenient because
the required expectation can be approximated using the core of the
sequential Monte Carlo update algorithm.

Assumption~\ref{assumption:qt} motivates a practical, approximate
solution to the ideal formulation of the dynamic programming problem
described in Section~\ref{subsec:1}. In the special case of linear
Gaussian state-space models, the vector $Q_t$ would form a sufficient statistic
since $\pi_t$ would be Gaussian,
and our approach would reduce to the standard Kalman filtering
procedure [see Chapter 12 of \citet{bd91} for details]. For the case of
nonlinear, non-Gaussian state-space models, such as the illustrative
one in this paper, $\pi_t$ is not summarized by a finite-dimensional
sufficient statistic. Assumption~\ref{assumption:qt} permits an
approximate solution to our American option pricing problem and, in
particular, motivates the conditional expectation in equation (\ref
{eq:mainrecursion}) by using $Q_t$ to summarize information about the
latent volatility using the past price history.
We next introduce two algorithms for pricing American-style options,
making use of the summary vectors $Q_t$ described in Section~\ref
{subsec:2}, and we illustrate
how our algorithms perform through a series of numerical experiments.


\section{Pricing algorithms} 
\label{sec:pricealg}

\subsection{A least-squares Monte Carlo based approach}
\label{subsec:montecarlo}

The popular least-squares Monte Carlo (LSM) algorithm of \citet{LongSch}
relies essentially on approximating the conditional expectation in
(\ref{eq:dprecursion}) by a regression function, in which one or
several recent values $S_t,S_{t-1},\dots$ are used as explanatory
variables. Since the conditional expectation is in fact (exactly) a
functional of the filtering distribution $\pi_t$, we might expect to
obtain some improvement by performing the regression using summary
features of $\pi_t$ as covariates instead.
This variant of the LSM algorithm,
describing the simulation component of our pricing methodology,
is stated in Algorithm~\ref{alg:smclsmone}.

%
\begin{algorithm}
\caption{Preliminary simulation of trajectories}
\label{alg:smclsmone}
\begin{algorithmic}[0]
\STATE\For{$n=1,\ldots,N$}

\quad\textit{Step 1.} Simulate a share price path $\{
S_0^{(n)},\ldots,S_T^{(n)} \}$ with $S_0^{(n)}=s_0$ from the
risk-neutral stochastic volatility model [equations
(\ref{eq:dtapprox1})--(\ref{eq:dtapprox2})].

\quad\textit{Step 2.} Apply Algorithm~\ref{alg:smc} (sequential Monte
Carlo algorithm) replacing $\{S_0,\ldots,S_T\}$ by the simulated
path $\{S_0^{(n)},\ldots,S_T^{(n)}\}$ to obtain approximations to
the filtering distributions $\{ \pi_t, ~t=0,\ldots,T \}$ for the
simulated path.

\quad\textit{Step 3.} Use the estimate of the filtering distribution
computed above in Step~2 to construct a summary vector, $Q_n$,
of $\pi_n(y_n)$ that stores key measures of the filtering
distribution such as the mean, standard deviation, skew, etc.

\quad\textit{Step 4.} Store the vector $(S_n, Q_n)$.

\EndFor

\textit{Repetition.} Repeat above steps to create $M$ independent
paths that contain information on the simulated share prices $S_t$ and
the summary vector $Q_t$ for all time points $t=0,1,2, \ldots, N$.
\end{algorithmic}
\end{algorithm}

\begin{remark*}
For the stochastic volatility model used in this analysis,
measures of center and spread will suffice to capture the key
features of the distribution. Therefore, our summary vector $Q_t =
(\mu_t, \zeta_t)$ in Algorithm~\ref{alg:smclsmone} describes the mean
and standard deviation of
the filtering distribution.
\end{remark*}

\begin{remark*}
The summary vector in Algorithm~\ref{alg:smclsmone} can include as many
key measures
of the filtering distribution $\pi_t(y_t)$ as needed to accurately
describe it. Other types of stochastic volatility models
may require additional measures that capture potential skewness,
kurtosis or modalities in the filtering distribution. One can learn
of the need for such measures by doing an empirical analysis on
historical data using Algorithm~\ref{alg:smc} to gain insights into
the behavior of the filtering distribution $\pi_t(y_t)$.
\end{remark*}

Next, we illustrate in Algorithm~\ref{alg:smclsmtwo} the implementation
of the LSM regression step using our summary vector $Q_t$ of $\pi_t$.

%
\begin{algorithm}
\caption{The least-squares Monte Carlo algorithm of
\protect\citet{LongSch} and the dynamic programming step}
\label{alg:smclsmtwo}

\begin{algorithmic}[0]
\STATE\textit{Initialization}

\qquad\textit{Sub-step A.} Run Algorithm~\ref{alg:smclsmone} to
obtain $M$ independent paths, where each path simulates realizations
of the share price $S_t$ and the summary vector $Q_t$ for time
points $t=1,2,\ldots,N$.

\qquad\textit{Sub-step B.} Compute the option price at $t=N$ along
each of the $M$ paths by evaluating the payoff function $g(S_T)$,
resulting in $M$ option values $\{\hat{u}^*_{T,1}, \ldots,
\hat{u}^*_{T,M}\}$.\vspace*{1pt}

\For{$t=N-1, N-2, \ldots, 1$}

\textit{Step 1.} Evaluate the exercise value $g(S_t^{(i)})$
for $i=1, \ldots, M$.\vspace*{1pt}

\textit{Step 2.} Compute basis functions of $S_t^{(i)}$ and
$Q_t^{(i)}$ for $i=1, \ldots, M$.

\textit{Step 3.} Approximate the hold value of the option at time
$t$ by
\[
E_{\mathrm{RN}}[\hat{u}^*_{t+1}(S_{t+1}, Q_{t+1})|S_t = s_t, Q_t=q_t] \approx
\sum_{k=1}^p \beta_{tk} \phi_k(S_t, Q_t),
\]
where the $\beta_{tk}$ are the coefficients of a regression (with $p$
explanatory variables) of the
discounted time $t+1$ American option prices,
$\hat{u}^*_{t+1}$, on basis functions $\phi_k$ of
$S_t$ and $Q_t$.

\textit{Step 4.} For $i=1, \ldots, M$, compute the exercise/hold
decision according to,
\begin{eqnarray*}
&&\hat{u}_t\bigl(S_t^{(i)}, Q_t^{(i)}, d_t\bigr) \\
&&\qquad= \cases{
g\bigl(S_t^{(i)}\bigr), &\quad $d_t = E$, \cr
E_{\mathrm{RN}}\bigl[\hat{u}^*_{t+1}(S_{t+1}, Q_{t+1})|S_t^{(i)}
= s_t^{(i)}, Q_t^{(i)}=q_t^{(i)}\bigr], &\quad $d_t = H$,
}
\end{eqnarray*}
\EndFor

Average the option values over
all $M$ paths to compute a Monte Carlo estimate and standard
error of the American option price.

\end{algorithmic}
\end{algorithm}

\begin{remark*} The regression in Step 3 of Algorithm~\ref
{alg:smclsmtwo} uses basis
functions of the share price $S_t$ and the summary vector $Q_t$ to
form the explanatory variables. We choose Laguerre functions as
basis functions. Our summary vector $Q_t = (\mu_t, \zeta_t)$
consists of the mean and standard deviation of the filtering
distribution $\pi_t(y_t)$. The design matrix used in the regression
at time point $k$ consists of the first two Laguerre functions in
$S_k$, $\mu_k$ and $\zeta_k$, and a few cross-terms of these
covariates. Specifically, our covariates used in the regression at
time $k$ (in addition to the intercept term) are
\begin{eqnarray*}
&
L_0(S_k),\qquad L_1(S_k),\qquad L_0(\mu_k),\qquad L_1(\mu_k),\qquad L_0(\zeta_k),\qquad L_1(\zeta_k),
&\\
&
L_0(S_k)*L_0(\mu_k),\qquad L_0(S_k)*L_0(\zeta_k),\qquad L_1(S_k)*L_1(\mu_k),
&\\
&L_1(S_k)*L_1(\zeta_k),\qquad L_0(\mu_k)*L_0(\zeta_k),\qquad L_1(\mu
_k)*L_1(\zeta_k),&
\end{eqnarray*}
where $L_0(x) = e^{-x/2}$ and $L_1(x) = e^{-x/2} (1-x)$ and, in\vspace*{1pt}
general, $L_n(x) = e^{-x/2} \frac{e^x}{n!} \frac{d^n}{dx^n} (x^n
e^{-x})$. Other choices of basis functions, such as Hermite\vspace*{1pt}
polynomials or Chebyshev polynomials, are also reasonable alternatives that
could be used to implement the least-squares Monte Carlo algorithm of
\citet{LongSch}.
\end{remark*}

\begin{remark*}
\citet{LongSch} actually adjust
Step 4 of Algorithm~\ref{alg:smclsmtwo} as follows:
\[
\hat{u}_t\bigl(S_t^{(i)}, Q_t^{(i)}, d_t\bigr) = \cases{
g(s_t), &\quad $d_t = E$, \cr
\hat{u}^*_{t+1}(S_{t+1}, Q_{t+1}), &\quad $d_t = H$,
}
\]
in order to avoid computing American option prices with a slight
upward bias due to Jensen's inequality; we follow their recommendation
and use this adjustment. They also suggest using only paths where
$g(S_t^{(i)}) > 0$ (i.e., the ``in-the-money'' paths) as a numerical
improvement.
However, we could use all paths since the convergence of the algorithm
also holds in this case
[see \citet{clp02}]. Therefore, we use all paths in our implementation of
Algorithm~\ref{alg:smclsmtwo}, as this produces similar results.
\end{remark*}

\subsection{A grid-based approach}
\label{subsec:grid}

We now present a grid-based algorithm for determining approximate
solutions to the dynamic programming problem. This \mbox{algorithm} is based
on a portion of the research work in \citet{RambharatPhD}. The approach
uses the
vectors $Q_t$ summarizing the
filtering distributions $\pi_t(\cdot)$ as arguments to the decision and
value functions. In contrast to the Monte Carlo based approach of
Section~\ref{subsec:montecarlo} where we directly use a summary vector
$Q_t$ in the LSM method, the grid-based technique requires a
distribution to approximate $\pi_t$. This distribution will typically be
parameterized by the summary vector when used to execute the
grid-based algorithm. In our illustrative pricing model, we use a
Gaussian distribution to approximate the filtering distribution, $\pi
_t$, at each time point $t$. \citet{KotechaDjuric03} provide some
motivation for using Gaussian particle filters, namely, Gaussian
approximations to the filtering distributions in nonlinear,
non-Gaussian state-space models. However, any distribution that
approximates $\pi_t$ reasonably well could be used for our purposes.
Such a choice will depend on the model and an empirical assessment of
$\pi_t$.

Since the steps of the
sequential Monte Carlo algorithm are designed to make the transition
from a specified $\pi_t$ to the corresponding distribution
$\pi_{t+1}$, we can combine a standard Monte Carlo simulation approach
with the use of Steps 1, 2 and 3 of Algorithm~\ref{alg:smc} to
compute the expectation. To be more specific, the next algorithm
computes the expectation on the right-hand side of
equation~(\ref{eq:mainrecursion}).

Algorithm~\ref{alg:ce} works by drawing pairs
$(s_{t+1}^{(i)},q_{t+1}^{(i)})$ from the conditional distribution of
$(S_{t+1},Q_{t+1})$, given $S_t=s_t, Q_t=q_t$, and using these to compute
a Monte Carlo estimator of the required conditional expectation.
Hence, we are able to evaluate $\hat{u}_t$ at various points, given
knowledge of $\hat{u}_{t+1}^*$,
and will thus form a key component of the backward induction step.
Note that this algorithm also relies on Assumption~\ref{assumption:qt}
in its use of the summary vector $Q_t$.

%
\begin{algorithm}
\caption{Estimation of conditional expectations in
equation~(\protect\ref{eq:mainrecursion})}
\label{alg:ce}
\begin{algorithmic}[0]
\STATE Draw values $\{y_t^{(i)}, i=1,\ldots,m\}$ independently
from a distribution chosen to be
consistent with a parameterization of the summary vector $Q_t=q_t$.
\vskip2mm
\For{$j=1,\ldots,n$}
\begin{itemize}
\item
\parbox[t]{11.85cm}{
{Draw $U \sim\operatorname{Unif}(1,\ldots,m)$. Then
draw $s_{t+1}^{(j)}$ from the conditional distribution of
$S_{t+1}$, given $S_t=s_t$ and $Y_t=y_t^{(U)}$.\vspace*{2pt}}
}
\item
\parbox[t]{11.85cm}{
Go through Steps 1 and 2 of Algorithm~\ref{alg:smc},
but in computing weights $\{w_{t+1}^{(i)},~i=1,\ldots,m\}$,
replace the actual log-return
$r_{t+1}$ by (the simulated log-return) $r_{t+1}^{(i)} = \log
(s_{t+1}^{(j)}/s_t)$.
}
\item
\parbox[t]{11.85cm}{
Go through Step 3 of Algorithm~\ref{alg:smc},
to obtain $\{y_{t+1}^{(1)},\ldots,y_{t+1}^{(m)}\}.$
}
\item
\parbox[t]{11.85cm}{
Compute $q_{t+1}^{(j)}$ as the appropriate summary vector.
}
\end{itemize}
\EndFor
\vskip2mm
\STATE
Compute the approximation
%
\begin{equation}\label{eq:mc01}
E_{\mathrm{RN}}[ \hat{u}_{t+1}^*(S_{t+1},Q_{t+1})
| S_t=s_t, Q_t=q_t]
 \simeq \frac{1}{n} \sum_{j=1}^n
\hat{u}_{t+1}^*\bigl(s_{t+1}^{(j)},q_{t+1}^{(j)}\bigr).
\end{equation}
\end{algorithmic}
\end{algorithm}

Since we will be interested in storing the functions
$\hat{u}_t^*(\cdot,\cdot)$ and
$\hat{d}_t^*(\cdot,\cdot)$,
we next introduce some additional notation. Let
%
\begin{equation}\label{eq:grid}
\mathcal{G} = \{ g_i \in\Rr^{d_q+1}, i=1,2,\ldots,G \}
\end{equation}
denote a collection of grid points in $\Rr^{d_q+1}$,
where $d_q$ denotes the dimensionality of $Q_t.$
These are points at which we will evaluate and
store the functions $\hat{u}_t^*$ and $\hat{d}_t^*$.
We will typically take
%
\begin{equation}
\mathcal{G} = \mathcal{G}_1 \times\mathcal{G}_2\times \cdots\times
\mathcal{G}_{d_q+1},
\end{equation}
where $\mathcal{G}_1$ is a grid of possible values for the share price,
and $\mathcal{G}_j$, $j>1$, is a grid of possible values for the $(j-1)$st
component of $Q_t$.

We state our grid-based pricing routine in Algorithm~\ref{alg:main}.
[This is a standard gridding approach to solving the dynamic
programming equations, as described, e.g.,
in \citet{BrockwellKadane}.]

%
\begin{algorithm}
\caption{Grid-based summary vector American option pricing algorithm}\label{alg:main}
\begin{algorithmic}[0]
\STATE\textit{Initialization.} For each $g \in\mathcal{G}$,
evaluate
\[
\hat{u}_T(g, d_T),\qquad \hat{d}_T^*(g),\quad \mbox{and}\quad
\hat{u}_T^*(g),
\]
using equations~(\ref{eq:terminal}), (\ref{eq:dstar}), and~(\ref{eq:ustar}).
Store the results.\vspace*{3pt}

\For{$t=T-1,T-2,\ldots,0$}

\quad\For{each $g \in\mathcal{G}$}

\qquad\parbox[t]{11.5cm}{
Evaluate
\[
\hat{u}_t(g, d_t),\qquad \hat{d}_t^*(g),\quad \mbox{and}\quad
\hat{u}_t^*(g),
\]
using equations~(\ref{eq:mainrecursion}),
~(\ref{eq:dstar}), and (\ref{eq:ustar}).
To evaluate the expectations in equation~(\ref{eq:mainrecursion}),
use Algorithm~\ref{alg:ce}.
Store the results.
}

\quad\EndFor

\EndFor

\vskip2mm
\STATE Evaluate the option price
%
\begin{equation}
\label{eq:optprice}
\mbox{price} = \hat{u}_0^*(s_0, q_0),
\end{equation}
where $s_0$ is an observed initial price and $q_0$ is an initial
(summary) measure
of the volatility process.
\end{algorithmic}
\end{algorithm}

\begin{remark*}
As is the case for the Monte Carlo based approach that we describe
in this paper, the grid-based scheme stated in Algorithm~\ref{alg:main}
also gives an option price which assumes that no information about
volatility is available at time $t=0$. In
the absence of such information, we just assume that the initial
log-volatility $Y_0$ can be modeled as coming from the limiting
distribution of the autoregressive process $\{Y_t\}$ and take $q_0$
as the summary measure of this distribution
(e.g., its mean and variance if the limiting distribution is Gaussian).
However, in most cases, it is possible to estimate log-volatility at
time $t=0$
using previous observations of the price process $\{S_t\}$.
In such cases, $q_0$ would represent the mean and conditional variance
of $Y_0$, given
``previous'' observations $S_{-1},S_{-2},\ldots,S_{-h}$ for some $h>0$.
These could be obtained in a straightforward manner by making use of
the sequential Monte Carlo estimation procedure described in
Algorithm~\ref{alg:smc} (appropriately modified so that
time $-h$ becomes time $0$). We could also base $q_0$ on a historical
volatility measure (i.e., the standard deviation of a few past observations).
\end{remark*}

\begin{remark*}
As with any quadrature-type approach, grid ranges must be chosen
with some care. In order to preserve quality of approximations
to the required expectations in Algorithm~\ref{alg:main}, it is necessary
for the ranges of the marginal grids $\mathcal{G}_j$ to contain
observed values of the respective quantities
with probability close to one.
Once the stochastic volatility model has been
fit, it is typically relatively easy to determine such ranges.
\end{remark*}

\begin{remark*}
The evaluation of $\hat{u}_t(g,d_t)$ in the previous algorithm
is performed making use of the Monte Carlo approximation given by
(\ref{eq:mc01}). Obviously the expression relies on knowledge
of $\hat{u}^*_{t+1}(\cdot,\cdot)$, but since we have only
evaluated $\hat{u}^*_{t+1}$ at grid points $g \in\mathcal{G}$, it
is necessary to interpolate in some manner. Strictly speaking,
one could simply choose the nearest grid point, and rely on sufficient
grid density to control error. However, inspection of the surface
suggests that
local linear approximations are more appropriate. Therefore, in
our implementations, we use linear interpolation between grid points.
\end{remark*}

\subsection{Numerical experiments}
\label{subsec:numerical}

The pricing algorithms outlined in Sections \ref{subsec:montecarlo}
and~\ref{subsec:grid} demonstrate how to price American-style options
in a latent stochastic volatility framework. These pricing algorithms
are computationally intensive, however, their value will depend on how
accurately they price American options in this partial observation
setting. We next illustrate the applicability of our pricing algorithms
through a series of numerical experiments. We assess the accuracy of
our valuation procedure by pricing American-style put options using the
following methods. (All methods use the current share price $S_t$ in
the LSM regression, however, the difference in each method is how
volatility is measured.)

\begin{itemize}

\item\textit{Method A (basic LSM).} This method simulates
asset prices $S_t$ according to the model (\ref{eq:dtapprox1})--(\ref{eq:dtapprox2}), however, the regression step in
the LSM algorithm uses a few past observations ($S_{t-1},
S_{t-2},\ldots$) as a measure of volatility in lieu of the summary
vector $Q_t$. This procedure is most similar to the traditional LSM
approach of \citet{LongSch}.

\item\textit{Method B (realized volatility).} This procedure
simulates asset prices $S_t$ according to the model (\ref
{eq:dtapprox1})--(\ref{eq:dtapprox2}), however, for
each time point $k$ ($k = 1,\ldots,N$) and for each path $l$ ($l =
1,\ldots,M$), we compute a measure of realized volatility
%
\begin{equation}\label{eq:rvol}
RV_{k,l} = \frac{1}{k} \sum_{j=1}^k R_{j,l}^2,
\end{equation}
where $R_{j,l}$ is the return at time $j$ along path $l$. The LSM
regression step then proceeds to use the realized volatility measure
$RV_{k,\cdot}$ at each time point $k$ as a measure of volatility.

\item\textit{Method C (MC/Grid).} This method uses the
algorithms described in Sections \ref{subsec:montecarlo} and
\ref{subsec:grid} to price American-style options in a latent stochastic
volatility framework either using (i) a pure simulation-based Monte
Carlo (MC) approach or (ii)~a Grid-based approach. Along with the
simulated share prices $S_t$, this approach makes extensive use of the
summary vectors $Q_t$ that capture key features of the volatility
filtering distribution $\pi_t$.

\item\textit{Method D (observable volatility).} This approach
simulates the asset prices $S_t$ and the volatility $\sigma_t$,
however, it assumes that \textit{both} asset price \textit{and}
volatility are observable. This is the full observation case that we
will use as a benchmark. Whichever of methods A, B or C is closest to
method D will be deemed the most accurate.
\end{itemize}

\begin{figure}

\includegraphics{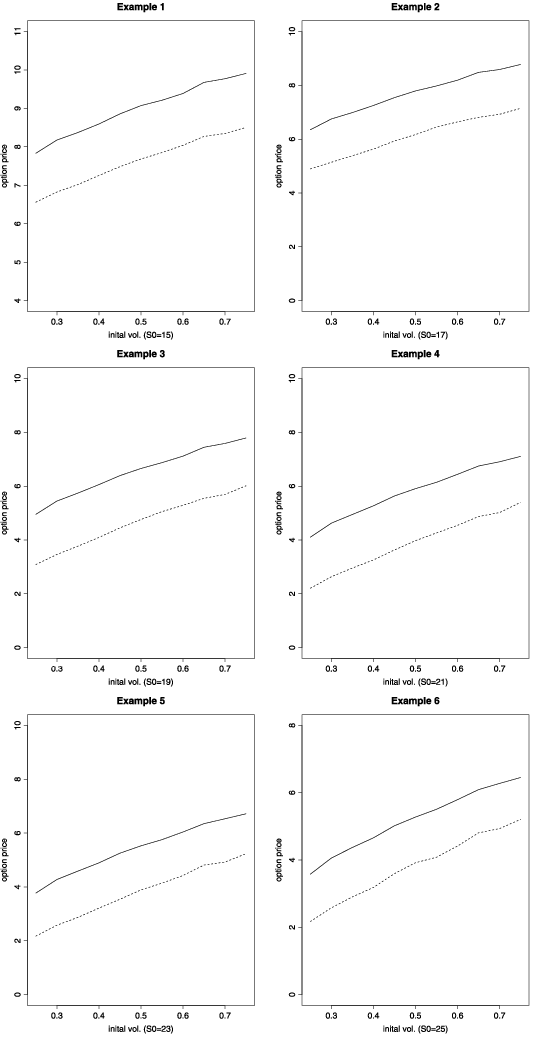}

\caption{A comparison of American put option pricing results
between methods \textup{A} (minimum observation case, dashed line) to method \textup{D}
(full observation case, solid line) for various parameter settings.}
\label{fig:partial}
\end{figure}

Figure~\ref{fig:partial} presents an illustrative example of the
difference in American put option prices between method A and method D
for several types of option contracts. One could think of this
illustration as reporting the difference in pricing results for two
extremes: the minimum observation case (method A or basic LSM) and the
full observation case (method D or observable volatility). This figure
uses parameter settings where stochastic volatility is prevalent. The
differences in option prices indeed show that stochastic volatility
matters when computing American option prices, especially when
volatility of volatility is high [i.e., when $\gamma$ in equation
(\ref
{eq:rnsvmodelall3}) is large].

%

In order to demonstrate the value of our proposed approach, method C
(MC/Grid), we must illustrate that it produces more accurate American
option pricing results than either of the simpler (and faster) methods
(A and B). We experimented with various model and option parameter
settings and found situations where simpler methods work just as well
as our approach. However, we also found
model/option parameter settings where our approach outperforms the
other pricing methods.
Table~\ref{tab:setup} describes the settings of our numerical
experiments. This table outlines selected values of the model
parameters and the American option pricing inputs for use in the
pricing algorithms ($K$ is the strike price, $T$ is the expiration in
days, $r$ is the interest rate, $S_0$ is the initial share price, and
$\sigma_0$ is a fixed initial volatility).

%
\begin{table}
\caption{Description of the stochastic volatility and American option
pricing inputs to the numerical experiments comparing methods \textup{A}, \textup{B}, \textup{C}
and \textup{D}. We set the number of particles $m=1000$ for~method \textup{C} and report
Macintosh OS X (V. 10.4.11) compute times for all cases}\label{tab:setup}
\begin{tabular}{@{}lcc@{}}
\hline
\textbf{Experiment no.} & \textbf{Parameters} $\bolds{(\rho, \alpha, \beta, \gamma, \lambda)}$ &
\textbf{Option inputs} $\bolds{(K, T, r, S_0, \sigma_0)}$ \\
\hline
1 & ($-$0.055, 3.30, $\log(0.55)$, 0.50, $-$0.10) & (23, 10, 0.055, 20,
0.50) \\
2 & ($-$0.035, 0.25, $\log(0.20)$, 2.10, $-$1.0) & (17, 20, 0.0255, 15,
0.35) \\
3 & ($-$0.09, 0.95, $\log(0.25)$, 3.95, $-$0.025) & (16, 14, 0.0325, 15,
0.30) \\
4 & ($-$0.01, 0.020, $\log(0.25)$, 2.95, $-$0.0215) & (27, 50, 0.03, 25,
0.50) \\
5 & ($-$0.03, 0.015, $\log(0.35)$, 3.00, $-$0.02) & (100, 50, 0.0225, 90,
0.35) \\
6 & ($-$0.017, 0.0195, $\log(0.70)$, 2.50, $-$0.0155) & (95, 55, 0.0325,
85, 0.75) \\
7 & ($-$0.075, 0.015, $\log(0.75)$, 6.25, 0.0) & (16, 17, 0.0325, 15,
0.35) \\
8 & ($-$0.025, 0.035, $\log(0.15)$, 5.075, $-$0.015) & (18, 15, 0.055, 20,
0.20) \\
9 & ($-$0.05, 0.025, $\log(0.25)$, 4.50, $-$0.015) & (19, 25, 0.025, 17,
0.35) \\
\hline
\end{tabular}
\end{table}

%
\begin{table}[b]
\tablewidth=295pt
\caption{American put option pricing results (and compute times) using
methods \textup{A} (basic LSM with past share prices) and \textup{B} (realized volatility
as an estimate for volatility)}\label{tab:lsm}
\begin{tabular*}{295pt}{@{\extracolsep{\fill}}lcc@{}}
\hline
\textbf{Experiment no.} & \textbf{A (basic LSM)} & \textbf{B (realized volatility)} \\
\hline
1 & 3.044 (0.00985) & 3.038 (0.00999) \\
Time (sec) & 18 & 12 \\
2 & 2.147 (0.00947) & 2.154 (0.00975) \\
Time (sec) & 15 & 16 \\
3 & 1.212 (0.00759) & 1.231 (0.00842) \\
Time (sec) & 16 & 14 \\
4 & 3.569 (0.0242) & 4.153 (0.0358) \\
Time (sec) & 37 & 39 \\
5 & 12.994 (0.0743) & 15.067 (0.120) \\
Time (sec) & 48 & 39 \\
6 & 18.623 (0.121) & 21.476 (0.168) \\
Time (sec) & 57 & 41 \\
7 & 1.588 (0.0120) & 1.843 (0.0186) \\
Time (sec) & 18 & 14 \\
8 & 0.0945 (0.00367) & 0.145 (0.00553) \\
Time (sec) & 11 & 15 \\
9 & 2.437 (0.0132) & 2.667 (0.0197) \\
Time (sec) & 21 & 21 \\
\hline
\end{tabular*}
\end{table}

%
\begin{table}
\tablewidth=226pt
\caption{American put option pricing results (and compute times) using
methods \textup{C} (MC/Grid) and \textup{D} (observable volatility)}\label{tab:smcfull}
\begin{tabular*}{226pt}{@{\extracolsep{\fill}}lcc@{}}
\hline
\textbf{Experiment no.} & \textbf{C (MC/Grid)} & \textbf{D (observable volatility)} \\
\hline
1 & 3.051 (0.0108) & 3.046 (0.00981) \\
Time (sec) & 162 & 15 \\
2 & 2.160 (0.00961) & 2.173 (0.00990) \\
Time (sec) & 310 & 15 \\
3 & 1.257 (0.00901) & 1.260 (0.00901) \\
Time (sec) & 218 & 12 \\
4 & 4.688 (0.0439) & 4.734 (0.0441) \\
Time (sec) & 760 & 37 \\
5 & 16.208 (0.138) & 16.331 (0.138) \\
Time (sec) & 815 & 39 \\
6 & 22.758 (0.184) & 22.996 (0.185) \\
Time (sec) & 900 & 46 \\
7 & 1.994 (0.0216) & 2.045 (0.0225) \\
Time (sec) & 266 & 17 \\
8 & 0.168 (0.00691) & 0.172 (0.00704) \\
Time (sec) & 232 & 15 \\
9 & 2.850 (0.0233) & 2.861 (0.0235) \\
Time (sec) & 416 & 20 \\
\hline
\end{tabular*}
\end{table}

Table~\ref{tab:lsm} reports the American option pricing results (and
standard errors) for methods A (basic LSM) and B (realized volatility)
and Table~\ref{tab:smcfull} reports the pricing results (and standard
errors) for methods C (MC/Grid) and D (observable volatility). The
computation times for all methods are also reported. For methods A, B
and D, we used $M=15\mbox{,}000$ LSM paths in the pricing exercise. When
running method C, we observed negligible differences between the
MC-based and grid-based approaches. Hence, we report the pricing
results (and standard errors) for method C using Algorithms~\ref
{alg:smclsmone} and~\ref{alg:smclsmtwo} with $M=15\mbox{,}000$ LSM paths.
(The MC-based approach also permits comparisons to the other approaches
in terms of standard errors.)

There are some notable observations to be made from the results of the
numerical experiments in
Tables~\ref{tab:lsm} and~\ref{tab:smcfull}. First, when the effect of
volatility is weak/moderate and the effect of mean reversion is
moderate/strong (experiments 1 and 2), all methods result in similar
American option pricing results. There are cases in the literature that
discuss fast mean reversion [see, e.g., \citet{fps00}], and in these
cases it would certainly make sense to use a faster pricing method.
However, in the situations where we experiment with dominant volatility
effects (experiments 3 through 9), although not often encountered in
the empirical stochastic volatility literature, but pertinent to
volatile market scenarios, method C comes closest to method D.
One should note that in cases of dominant volatility method B does
better than method A, as it uses a more accurate measure of volatility.
Method C, however, which actually makes use of the filtering
distributions $\pi_t$, comes within standard error of method D
(observable volatility case). Additionally, we observed when stochastic
volatility is dominant and the American option has a long maturity and
is at/in-the-money, the difference in pricing results is more pronounced.

Method C (MC/Grid) is a computationally intensive approach, although
this feature is shared by many Monte Carlo and grid-based techniques.
Clearly, the proposed pricing algorithm using method C is not
competitive in terms of computation time. The simpler methods (A and B)
are much faster and also accurate under strong mean reversion and weak
stochastic volatility. It should be observed, however, that the pricing
accuracy is much higher using method C, as it is always within standard
error of method D (observable volatility). The accuracy of method C
holds in all model/option parameter settings. One does not require
special cases (high mean reversion, low volatility) or special option
contract features (long/short maturity, in/at/out-of-the money) for our
approach to be competitive in terms of accuracy.
We also ascertain the robustness of our approach by computing (and
storing) the exercise rule from each of the four methods (A through D).
We use the rule to revalue the American put options on an independent,
common set of paths. The results are described in Tables~\ref
{tab:lsmrobust} and~\ref{tab:smcfullrobust}; note that the respective
pricing results are similar to those stated in Tables~\ref{tab:lsm}
and~\ref{tab:smcfull}, hence leading to similar conclusions.

%
\begin{table}[b]
\tablewidth=240pt
\caption{American put option pricing results (and compute times) using
the exercise rule from methods \textup{A} and \textup{B} on an independent, common set of
paths. The results are comparable to those given in Table~\textup{\protect\ref{tab:lsm}}}\label{tab:lsmrobust}
\begin{tabular*}{240pt}{@{\extracolsep{\fill}}lcc@{}}
\hline
\textbf{Experiment no.} & \textbf{A (basic LSM)} & \textbf{B (realized volatility)} \\
\hline
1 & 3.045 (0.00993) & 3.050 (0.0101) \\
Time (sec) & 31 & 16 \\
2 & 2.128 (0.00943) & 2.141 (0.00991) \\
Time (sec) & 23 & 17 \\
3 & 1.213 (0.00753) & 1.239 (0.00856) \\
Time (sec) & 21 & 20 \\
4 & 3.540 (0.0242) & 4.162 (0.0365) \\
Time (sec) & 59 & 57 \\
5 & 13.043 (0.0736) & 15.035 (0.120) \\
Time (sec) & 72 & 58 \\
6 & 18.225 (0.122) & 20.934 (0.168) \\
Time (sec) & 65 & 70 \\
7 & 1.568 (0.0115) & 1.809 (0.0180) \\
Time (sec) & 25 & 23 \\
8 & 0.106 (0.00428) & 0.156 (0.00590) \\
Time (sec) & 17 & 20 \\
9 & 2.434 (0.0131) & 2.669 (0.0198) \\
Time (sec) & 30 & 32 \\
\hline
\end{tabular*}
\end{table}

%
\begin{table}
\tablewidth=240pt
\caption{American put option pricing results (and compute times) using
the exercise rule from methods \textup{C} and \textup{D} on an independent, common set of
paths. The results are comparable to those given in Table~\textup{\protect\ref{tab:smcfull}}}\label{tab:smcfullrobust}
\begin{tabular*}{240pt}{@{\extracolsep{\fill}}lcc@{}}
\hline
\textbf{Experiment no.} & \textbf{C (MC/Grid)} & \textbf{D (observable volatility)} \\
\hline
1 & 3.050 (0.0107) & 3.050 (0.00977) \\
Time (sec) & 193 & 20 \\
2 & 2.145 (0.00966) & 2.151 (0.00990) \\
Time (sec) & 302 & 23 \\
3 & 1.271 (0.00913) & 1.276 (0.00916) \\
Time (sec) & 253 & 16 \\
4 & 4.735 (0.0441) & 4.813 (0.0448) \\
Time (sec) & 747 & 62 \\
5 & 16.092 (0.137) & 16.185 (0.139) \\
Time (sec) & 740 & 63 \\
6 & 22.380 (0.184) & 22.646 (0.186) \\
Time (sec) & 885 & 80 \\
7 & 1.965 (0.0213) & 2.003 (0.0220) \\
Time (sec) & 274 & 21 \\
8 & 0.186 (0.00747) & 0.196 (0.00785) \\
Time (sec) & 228 & 18 \\
9 & 2.868 (0.0236) & 2.894 (0.0238) \\
Time (sec) & 378 & 30 \\
\hline
\end{tabular*}
\end{table}

The differences in pricing results, as noted in Tables~\ref{tab:lsm}
and~\ref{tab:smcfull} (or Tables~\ref{tab:lsmrobust} and~\ref
{tab:smcfullrobust}), are relevant when thinking about how some option
transactions take place in practice. For example, on the American Stock
Exchange (AMEX), American-style options have a minimum trade size of
one contract, with each contract representing 100 shares of an
underlying equity.\footnote{Source: \url{http://www.amex.com}.}
Hence, the discrepancies between methods A and B and our proposed
method C could be magnified under such trading scenarios.
Our proposed approach would be especially useful when constructing risk
management strategies during volatile periods in the market.

We also repeated the numerical experiments in Table~\ref{tab:setup}
using a first-order Euler discretization of the model as opposed to
exact simulation. The Euler-based results are reported in Tables~\ref
{tab:lsmeuler} and~\ref{tab:smcfulleuler} of Section~\ref
{subsec:euler}. Upon inspection, one observes that the corresponding
results for each of methods A through D are virtually identical. Thus,
if a model does not permit exact simulation, a numerical procedure such
as the first-order Euler scheme or any other scheme [see, e.g., \citet{KloedenPlaten00}] should suffice for the purposes of executing our
pricing algorithm.

We now illustrate a statistical application of our proposed pricing
methodology. We first demonstrate how to estimate model parameters.
Next, we make inference, using observed American put option prices, on
the market price of volatility risk. Since method C outperforms either
A or B in all model/option settings, we will use it as our primary tool
for statistical analysis.


\section{Statistical estimation methodology} 
\label{sec:statmethod}

\subsection{Model parameter estimation}
\label{subsec:modelparameters}

The stochastic volatility model, outlined in equations (\ref
{eq:rnsvmodelall1}), (\ref{eq:rnsvmodelall2}) and~(\ref
{eq:rnsvmodelall3}), has been analyzed extensively in previous work,
namely, \citeauthor{Jacquieretal} (\citeyear{Jacquieretal}, \citeyear{jpr04}) and \citet{junyu05}. These
authors analyze the model under the statistical (or ``real-world'')
measure as described in a Section~\ref{sec:svmodel} footnote. Model
estimation of share prices under the real-world measure would only
require data on price history.
Estimation of model parameters in a risk neutral setting, however, is a
bit more involved as we need both share price data on the equity as
well as option data. \citet{pan02} illustrates how to use both share and
option price data to jointly estimate parameters under the real-world
measure and risk-neutral measure. \citet{Eraker} also implements joint
estimation methodology for share and
option price data. However, both \citet{pan02} and \citet{Eraker} only
deal with the case of European-style options.

We propose a two-step procedure to estimate our illustrative stochastic
volatility model in an American-style derivative pricing framework. In
the first step, we use share price data to estimate the statistical
model parameters.\vspace*{1pt}
Define the parameter vector\footnote{Strictly speaking, estimation
using only share prices (i.e., under the physical measure) involves the
physical drift rate in the parameter vector. However, since we do not
use the physical drift rate in risk-neutral pricing, we do not present
its estimation results.}
\[
\theta= (\rho, \alpha, \beta, \gamma).
\]
Conditional on estimated model parameter values, we then estimate the
market price of volatility risk parameter $\lambda$. Estimation of
$\lambda$ requires data on both share and American option prices.
Although it would be comprehensive to do a joint statistical analysis
of both share and option prices, this problem is quite formidable in
the American option valuation setting. (The full joint estimation
problem is left for future analysis.)
We adopt a Bayesian approach to parameter estimation. The first
objective is to estimate the posterior distribution of $\theta$,
%
\begin{equation}\label{eq:thetapost}
p(\theta|r_1,\ldots,r_n) = \frac{p(r_1,\ldots,r_n|\theta) \cdot
p(\theta)}{\int p(r_1,\ldots,r_n|\theta) \cdot p(\theta)\, d\theta}.
\end{equation}
Since $\rho\in[-1,1]$, $\alpha> 0$, $\beta\in\mathbb{R}$, and
$\gamma> 0$, we reparameterize a sub-component of the vector $\theta$
so that each component will have its domain in $\mathbb{R}$. This
reparameterization facilitates exploration of the parameter space. Let
us define
\[
\tilde{\rho} = \tan\biggl(\frac{\rho\pi}{2} \biggr),\qquad \tilde{\alpha} = \log
(\alpha
),\quad \mbox{and}\quad \tilde{\gamma} = \log(\gamma).
\]
We assign independent standard normal priors to the components of the
(reparameterized) vector
\[
\tilde{\theta} = [\tilde{\rho}, \tilde{\alpha}, \beta, \tilde
{\gamma} ].
\]
The next step is the evaluation of the likelihood (or log-likelihood)
for $\tilde{\theta}$, however, an analytical expression for the
likelihood is not available in closed-form. We employ Kitagawa's
algorithm [see, e.g., \citeauthor{kita87} (\citeyear{kita87,kita96}) and \citet{KitagawaSato} and the references therein]
to estimate the log-likelihood of our model for share prices under the
real-world measure. Kitagawa's algorithm, used to compute the
log-likelihood for nonlinear, non-Gaussian state-space models, employs
the fundamental principles of particle-filtering. The essence of
Kitagawa's approach uses the weights described in equation (\ref
{eq:basicweights}) of Algorithm~\ref{alg:smc} to provide a Monte Carlo
based approximation to the log-likelihood. The details of this
log-likelihood approximation are available in \citet{KitagawaSato}. We
provide Kitagawa's log-likelihood estimation approach for nonlinear,
non-Gaussian state-space models in Algorithm~\ref{alg:kitagawa}.

%
\begin{algorithm}[t]
\caption{\protect\citet{kita87} log-likelihood approximation}
\label{alg:kitagawa}
\begin{algorithmic}[0]

\STATE\textit{Initialization 1.} Input a proposed value of the parameter
vector $\theta$ for which the log-likelihood value is required.

\vskip2mm

\STATE\textit{Initialization 2.} Choose a number of ``particles'' $m>0$.
Draw a sample $\{\tilde{y}_0^{(1)},\ldots,\tilde{y}_0^{(m)}\}$ from
the distribution of $Y_0$ [see equation~(\ref{eq:initydn})].

\vskip2mm

\STATE\textit{Step 1.} Cycle through the (i) forward simulation, (ii)
weighting, and (iii) resampling steps of Algorithm \ref{alg:smc}
ensuring that the weights, $w_t^{(i)}$, $i=1, \ldots, m$, from equation
(\ref{eq:basicweights}) are stored for $t = 1, \ldots, T$.

\vskip2mm

\STATE\textit{Step 2.} Approximate the log-likelihood, $l(\theta)$, by
%
\begin{equation}
\label{eq:loglike}
l(\theta) \approx\sum_{t=1}^T \frac{1}{m} \sum_{i=1}^m \log\bigl(
w_t^{(i)} \bigr)
\end{equation}
\end{algorithmic}
\end{algorithm}

\begin{remark*}
The approximation to the log-likelihood value associated with a
particular parameter value in equation (\ref{eq:loglike}) of
Algorithm~\ref{alg:kitagawa} becomes more accurate as the number of
particles $m$ gets large. (We used $m=500$ in our estimation exercise.)
When resampling using the weights in equation (\ref{eq:basicweights}),
we sometimes work with the shifted log-weights as this leads to
improved sampling efficiency.
\end{remark*}

Once we obtain an estimate of the log-likelihood for the model
parameters, we then combine it with the
(log) priors and use a random-walk Metropolis algorithm to estimate the
(log) posterior distribution of $\tilde{\theta}$ (or, equivalently,
$\theta$). That is, we estimate the posterior distribution for $\tilde
{\theta}$ and then transform back to the original scale to calculate
the posterior distribution of $\theta$ as stated in equation (\ref
{eq:thetapost}). Our Markov chain Monte Carlo (MCMC) algorithm utilizes
a Gaussian proposal density in order to facilitate the estimation of
the posterior distribution in the $\tilde{\theta}$ parameter space. The
details of our MCMC sampler are described in Algorithm~\ref{alg:mcmc},
which is found in Section~\ref{subsec:mcmcvolrisk}.

Given the estimates of the model parameters under the real-world
measure using the share prices, we now describe how to use both share
and option price data to estimate the market price of volatility risk.
We work with a summary measure of $p(\theta|r_1, \ldots, r_n)$, namely,
the posterior mean, although one could use another measure such as the
posterior median. The estimation of the market price of volatility risk
will be computed conditional on this posterior summary measure.
We implement this estimation exercise using the algorithms outlined in
Section~\ref{sec:pricealg}.

\subsection{Volatility risk estimation}\label{subsec:volrisk}

The estimation of the market price of volatility risk, $\lambda$, in
equation (\ref{eq:rnsvmodelall3}) presents a computational challenge,
particularly in the American option valuation framework. We aim to use
the algorithms described in this paper to propose methodology that will
facilitate statistical inference of $\lambda$ in the American option
pricing setting. To the best of our knowledge, this is the first
analysis to compute and make posterior inference on the volatility risk
premium for American-style (early-exercise) options. In many
applications of option-pricing in a stochastic volatility framework, it
is often the case that $\lambda$ is set to a prespecified value [see,
e.g., \citet{hes93}, \citet{hw87} and \citet{pastrentou}].
One convenient approach is to set $\lambda= 0$. This is known as the
``minimal martingale measure'' in some strands of the option-pricing
literature [\citet{mr98}].

Both \citet{pan02} and \citet{Eraker} estimate stochastic volatility
model parameters, including the market price of volatility risk, for
the case of European options. However, the early-exercise feature of
American-style options adds further complexities to the estimation
problem. One method to estimate the market price of volatility risk for
American-style options would be to set up the following nonlinear
regression model [similar in spirit to what \citet{Eraker} does in his
analysis of European options]. Let
%
\begin{equation}
\label{eq:optionmodel}
U_i = P_{\theta^*}^i(\lambda) + \epsilon_i,
\end{equation}
where $U_i$ $(i=1,\ldots,L)$ is the observed American option price,
$P_{\theta^*}^i(\lambda)$ is the model predicted American option price
conditional on the mean, $\theta^*$, of the posterior distribution in
equation (\ref{eq:thetapost}), and $\lambda$ is the market price of
volatility risk.\footnote{Although it is not explicitly stated in
equation (\ref{eq:optionmodel}), both $\theta^*$ and $P_{\theta
^*}^i(\lambda)$ depend on the share price data $S_t$ or, equivalently,
the returns data $R_t$.}
$P_{\theta^*}^i(\lambda)$ is computed using one of our proposed pricing
algorithms in Section~\ref{sec:pricealg}. The error term, $\epsilon_i$,
is assumed to be an independent sequence of $N(0, \sigma^2)$ random variables.

The next step is to find the optimal value for $\lambda$ which we will
denote by $\lambda^*$. One approach would be to minimize the
sum-of-squared errors, $S(\lambda)$, where
%
\begin{eqnarray}
\label{eq:sse}
S(\lambda) &=& \sum_{i=1}^L \bigl(U_i - P_{\theta^*}^i(\lambda) \bigr)^2\quad \mbox{and}
\\
\label{eq:argminlambda}
\lambda^* &=& \mathop{\arg\min}_{\lambda} S(\lambda).
\end{eqnarray}
As noted in \citet{sebwild03}, the minimum value of $S(\lambda)$
corresponds to the least-squares estimate of the nonlinear regression
model in equation (\ref{eq:optionmodel}). One can also show that the
least-squares estimate is equivalent to the maximum likelihood estimate
(MLE). Optimizing $S(\lambda)$, although computationally demanding, is
feasible. We again adopt a Bayesian approach, outlined more generally
in \citet{sebwild03}, to solve this optimization problem.
First, we start with the (Gaussian) likelihood for the model in
equation (\ref{eq:optionmodel}) which is given by
\begin{eqnarray}
\label{eq:optionlike}
p(u_1, \ldots, u_L|\lambda, \sigma^2) &=& (2\pi\sigma^2 )^{-L/2} \sum_{i=1}^L \exp \biggl\{-\frac{1}{2 \sigma^2} \bigl(u_i -
P_{\theta^*}^i(\lambda) \bigr)^2 \biggr\} \nonumber\\[-8pt]\\[-8pt]
&=& (2\pi\sigma^2 )^{-L/2} \exp \biggl\{-\frac{1}{2 \sigma^2}
S(\lambda) \biggr\},\nonumber
\end{eqnarray}
where the second equality in the likelihood formulation follows from
equation~(\ref{eq:sse}).
As suggested in \citet{sebwild03}, if we use the following (improper)
prior distribution over $(\lambda, \sigma^2)$,
%
\begin{equation}\label{eq:priors}
p(\lambda, \sigma^2) \propto\frac{1}{\sigma^2},
\end{equation}
it follows that the posterior distribution for $(\lambda, \sigma^2)$
is, up to a constant of proportionality,\footnote{For simplicity, we
will suppress dependence on $R_t$ and $\theta^*$ in the calculation for
the posterior distribution of $\lambda$.}
%
\begin{equation}
\label{eq:jointposterior}
p(\lambda, \sigma^2|u_1, \ldots, u_L) \propto(\sigma^2 )^{- (
L/2+1 )} \exp \biggl\{-\frac{S(\lambda)}{2 \sigma^2} \biggr\}.
\end{equation}
Recognizing the kernel of the inverse gamma distribution for $\sigma^2$
in equation (\ref{eq:jointposterior}), namely, $\operatorname{IG} (\frac{L}{2},
\frac
{S(\lambda)}{2} )$, we can integrate out $\sigma^2$ to conclude that
%
\begin{equation}
\label{eq:lambdapost}
p(\lambda|u_1, \ldots, u_L) \propto\frac{\Gamma(L/2)}{ ( S(\lambda)/2
)^{L/2}} \propto(S(\lambda) )^{-L/2}.
\end{equation}
Therefore, we have shown in equation (\ref{eq:lambdapost}) that, with
the choice of prior
in equation (\ref{eq:priors}), the posterior distribution of the market
price of volatility risk $\lambda$ is proportional to $S(\lambda
)^{-L/2}$. If we maximize this posterior distribution, it is equivalent
to minimizing $S(\lambda)$, and hence, the result would be the same as
the least-squares estimate or the MLE.

One approach to approximating the posterior distribution in equation
(\ref{eq:lambdapost}) is to use an MCMC based procedure. However, in
the context of American option valuation, the early-exercise feature
would present a major computational challenge when evaluating
$S(\lambda)$. A more feasible approach is to use a result that concerns the
approximate normality of the posterior distribution close to the
posterior mode [see Chapter 2 of \citet{sebwild03} for a discussion in
the context of nonlinear regression]. If we denote the posterior mode
of equation (\ref{eq:lambdapost}) by $\lambda^*$, then under suitable
regularity conditions [see Chapter 7 of \citet{Schervish95}], the
posterior distribution of $\lambda$ near $\lambda^*$ can be
approximated as a normal distribution with mean $\lambda^*$ and
variance $V^*$. In particular,
%
\begin{equation}
\label{eq:lambdanormal}
\lambda|u_1, \ldots, u_L \sim N (\lambda^*, V^* ),
\end{equation}
where
%
\begin{equation}
\label{eq:lambdapostvar}
\frac{1}{V^*} = -\frac{d^2 \log (p(\lambda|u_1, \ldots, u_L)
)}{d\lambda^2} \bigg|_{\lambda= \lambda^*} = -\frac{d^2 \log
[(S(\lambda))^{-L/2} ]}{d\lambda^2} \bigg|_{\lambda= \lambda^*}.
\end{equation}
Algorithm~\ref{alg:volrisk} in Section~\ref{subsec:lambdapost} of the
\hyperref[sec:appendix]{Appendix} illustrates how to estimate the parameters of the normal
distribution in equation (\ref{eq:lambdanormal}). We use a grid-search
to find the posterior mode, $\lambda^*$, and then we estimate the
derivative expression in equation (\ref{eq:lambdapostvar}) using
numerical approximation techniques (namely, central differences)
described in, for instance, \citet{whd95}. We next report the results of
a data-analytic study of our American option valuation approach using
the aforementioned algorithms on three equities.


\section{Empirical analysis} 
\label{sec:dataexercise}
Our empirical analysis uses the algorithms outlined in Section~\ref
{sec:pricealg} together with historical share prices and American put
option data. A~reference to our computing code and data sets is given
in \citet{RamBrockSupp}.

\subsection{Data description}
\label{subsec:data}

\begin{figure}

\includegraphics{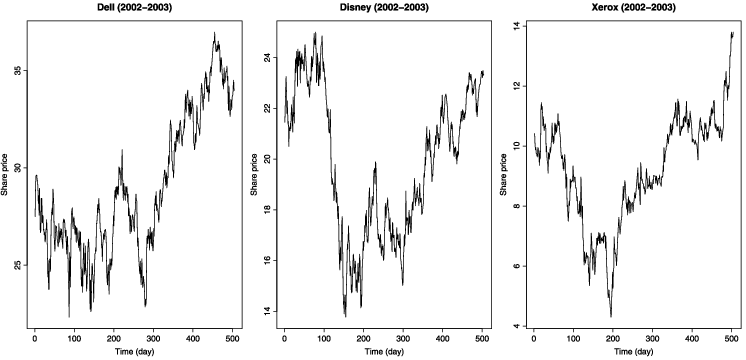}

\caption{A time series plot of the share prices of Dell, Disney and
Xerox over the period Jan. 2nd, 2002 to Dec. 31st, 2003.}
\label{fig:equities}
\end{figure}

We obtain observed market data on equity prices as well as
American-style put options on these underlying equities. We gather
share price data on three equities: Dell Inc., The Walt Disney Company
and Xerox Corporation. The share price data are sourced from the
Wharton Research Data Services (WRDS).\footnote{Access to the WRDS
database was granted through Professor Duane Seppi and the Tepper
School of Business at Carnegie Mellon University.}
The share price data represent two periods: (i) a historical period
from Jan. 2nd, 2002 to Dec. 31st, 2003, and (ii) a valuation period
from Jan. 2nd, 2004 to Jan. 30th, 2004 (the first 20 trading days of
2004). The historical share price data will be used for model parameter
estimation and the trading share price data will be used in the option
valuation analysis. The option price data sets are obtained for the
period spanning the first 20 trading days of 2004. The data on American
put options are extracted from the website of the American Stock
Exchange (AMEX). We also use the LIBOR rates from Jan. 2004 (1-month
and 3-month rates were around 0.011, and 6-month rates were around
0.012), obtained from $\mbox{Bloomberg}^{\textrm{\textregistered}}$,
for the value of the risk-free rate $r$. A plot of the share prices of
the three equities over the historical period appears in Figure~\ref
{fig:equities}. Additionally, Tables~\ref{tab:datashares} and~\ref
{tab:dataoption} summarize some features of the share and option price
data; note that most options are at-the-money and their maturities
range from short to long.

%

%
\begin{table}[b]
\tablewidth=260pt
\caption{Description of the equity share prices: the share price range
(in dollars) and the share price mean (and standard deviation) for the
historical period Jan. 2nd, 2002 to Dec. 31st, 2003 (estimation period)}\label{tab:datashares}
\begin{tabular*}{260pt}{@{\extracolsep{\fill}}lcc@{}}
\hline
\textbf{Equity} & \textbf{Historical range} & \textbf{Historical mean (sd)} \\
\hline
Dell & 22.33--36.98 & 28.95 (3.57) \\
Walt Disney & 13.77--25.00 & 19.83 (2.87) \\
Xerox & \hphantom{0}4.30--13.80 & \hphantom{0}9.18 (1.82) \\
\hline
\end{tabular*}
\end{table}

%
\begin{table}
\caption{Description of the American put options: the number of options
in our data set ($L$), the maturity (in days), the strike price (in
dollars), the share price range, and the share price mean (and standard
deviation) for the Jan. 2004 valuation period}\label{tab:dataoption}
\begin{tabular*}{\textwidth}{@{\extracolsep{\fill}}lccccc@{}}
\hline
\textbf{Equity} & $\bolds{L}$ & \textbf{Maturity} & \textbf{Strike} & \textbf{Share price range} & \textbf{Share price mean (sd)} \\
\hline
Dell & 120 & 15--98\hphantom{0} & 32.50--37.50 & 33.44--35.97 & 34.89 (0.65) \\
Walt Disney & \hphantom{0}60 & 15--135 & 25.00 & 23.67--24.96 & 24.45 (0.40) \\
Xerox & \hphantom{0}60 & 15--135 & 14.00 & 13.39--15.15 & 13.99 (0.48) \\ \hline
\end{tabular*}
\end{table}

\begin{figure}

\includegraphics{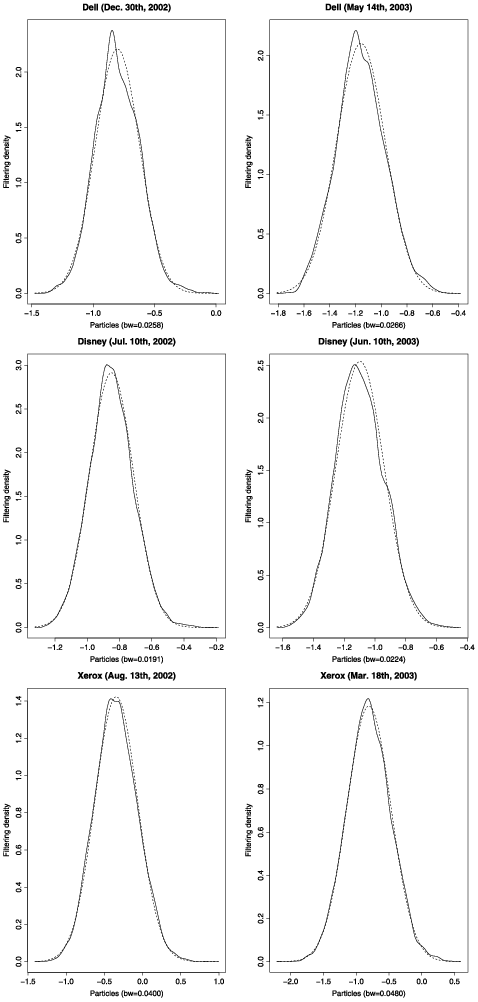}

\caption{Examples of the sequential Monte Carlo filtering
distributions, $\pi_t$, as defined in Section~\protect\ref{subsec:1} for each
of the three equities at a few selected dates in the estimation period
2002--2003. The solid lines are kernel-smoothed sequential Monte Carlo
estimates of $\pi_t$ and the dashed lines are the Gaussian approximations.}
\label{fig:pit}
\end{figure}

Figure~\ref{fig:pit} depicts $\pi_t$ along with Gaussian approximations
using the output
of Algorithm~\ref{alg:smc} for the
three equities in our analysis. We choose two time points for each
equity for our graphical illustrations, however, it should be noted
that results are similar for other time points. We construct the
summary vectors $Q_t$ based on these distributions. The grid-based
Algorithms \ref{alg:ce} and \ref{alg:main} would, for instance, use a
Gaussian distribution to approximate these filtering distributions, as
this appears to provide an adequate fit. Note that for other choices of
stochastic volatility models, a different (possibly non-Gaussian)
distribution may be suitable as an approximation to the filtering distribution.

\subsection{Posterior summaries}
\label{subsec:results}

We report the posterior means and 95\% credible intervals for the
parameter vector $\theta= (\rho, \alpha, \beta, \gamma)$ for each of
the three equities in Table~\ref{tab:thetaresults}. These are the
results from execution of Algorithm~\ref{alg:mcmc}. Inspection of
$\alpha$ and $\gamma$ shows that the volatility process for all
equities exhibit noticeable signs of mean reversion and stochastic
volatility, respectively. The results for $\beta$, the overall level of
the volatility process around which its mean reverts, are also
reported. Observe, as well, that the results for both Dell and Disney
show strong signs of the leverage effect between share prices and their
volatility. This is evidenced by the negative values for $\rho$ and the
fact that the 95\% credible intervals do not span zero. On the other
hand, the results for Xerox are not as conclusive, as the 95\% credible
interval for $\rho$ spans zero.

Conditional on the posterior means reported in Table~\ref
{tab:thetaresults}, we next estimate the posterior distribution of the
market price of volatility risk parameter, $\lambda$, for each equity.
This is facilitated by the implementation of Algorithm~\ref
{alg:volrisk}. Essentially, for each equity, we find the posterior mode
[i.e., the maximum value of the expression in equation (\ref
{eq:lambdapost})] and then we implement the analysis described in
Section~\ref{subsec:volrisk}.
Our results from the volatility risk estimation are reported in
Table~\ref{tab:lambdaresults}. We explain the full details of the
numerical computations in Section~\ref{subsec:lambdapost}.

%
\begin{table}
\caption{Bayesian posterior means and 95\% credible intervals (CI) for
the parameters of the
stochastic volatility model in
equations (\protect\ref{eq:rnsvmodelall1})--(\protect\ref{eq:rnsvmodelall3}) by equity}\label{tab:thetaresults}
\begin{tabular*}{\textwidth}{@{\extracolsep{\fill}}lcccc@{}}
\hline
\textbf{Equity} & $\bolds{\rho}$ & $\bolds{\alpha}$ & $\bolds{\beta}$ & $\bolds{\gamma}$ \\
\hline
Dell & $-$0.673 & 1.830 & $-$1.087 & 1.081 \\
95\% CI & ($-$0.767, $-$0.402) & (0.267, 5.398) & ($-$2.157, $-$0.316) &
(0.674, 1.695) \\
Disney & $-$0.612 & 0.363 & $-$1.379 & 0.686 \\
95\% CI & ($-$0.761, $-$0.259) & (0.0194, 1.805) & ($-$2.970, $-$0.288) &
(0.426, 1.080) \\
Xerox & 0.198 & 26.726 & $-$0.812 & 3.494 \\
95\% CI & ($-$0.0215, 0.419) & (8.036, 54.328) & ($-$1.030, $-$0.603) &
(2.119, 5.214) \\
\hline
\end{tabular*}
\end{table}

%
\begin{table}[b]
\tablewidth=250pt
\caption{Parameters of the normal approximation to the posterior
distribution of $\lambda$, the market price of volatility risk, as well
95\% credible intervals}\label{tab:lambdaresults}
\begin{tabular*}{250pt}{@{\extracolsep{\fill}}ld{3.3}cc@{}}
\hline
\textbf{Equity} & \multicolumn{1}{c}{$\bolds{\lambda^*}$} & $\bolds{V^*}$ & \textbf{95\% credible interval} \\
\hline
Dell & -6.350 & 0.00266 & $(-6.451, -6.249)$ \\
Disney & -10.850 & 0.00750 & $(-11.020, -10.680)$ \\
Xerox & -0.700 & 0.00815 & $(-0.877, -0.523)$ \\
\hline
\end{tabular*}
\end{table}

Based on the posterior analysis of $\lambda$, all three equities show
evidence of a negative value for this parameter. Observe that the
(approximate) 95\% credible intervals are negative and do not span
zero. This is consistent with results reported in the literature on the
market price of volatility risk [see e.g., \citeauthor{bk03a} (\citeyear{bk03a,bk03b})]. In
these studies, it is explained that the negative volatility risk
premium signals that investors are willing to pay a premium for
``downside protection'' (or adverse movements in share prices due to
stochastic volatility). This results because a negative value for
$\lambda$ implies a higher volatility mean reversion level [see
equation (\ref{eq:rnsvmodelall3})] and, therefore, most likely a higher
volatility and option price. It is especially during these adverse
movements (or volatile market activity) that our pricing methodology
for American-style options is most pertinent.
Furthermore, the magnitude of our results for the market price of
volatility risk in this small empirical analysis are in agreement with
studies that analyze a larger set of individual equities as well as
index options [\citet{bk03b}]. However, these earlier strands of
empirical research do not analyze options with early-exercise features.

Additionally, in Table~\ref{tab:lambdasse}, we report the
sum-of-squared errors, $S(\lambda$), when $\lambda$ equals $\lambda^*$
and when $\lambda=0$. Clearly, the model-predicted American option
prices better match market data for the optimized (nonzero) $\lambda$
value. This casts some evidence in favor of a nonzero volatility risk premium.
It is also interesting to note that the market price of volatility risk
parameter $\lambda$ does not appear to be the same across all equities.
Thus, if one had a portfolio of equities, it may be interesting to
understand the differences in their volatility risk premiums. One
potential reason for this difference across equities is that the market
price of volatility risk may be comprised of two components: (i) a
market component, which we may expect to be constant across equities,
and (ii) an idiosyncratic component, which may well be the fundamental
source of the differences in the estimates of $\lambda$ across equities
(Nikunj Kapadia, personal communication).
Further analyses call for more elaborate specifications of $\lambda$
and additional study of its underlying components. We could, for
instance, model $\lambda$ as a time-varying function or even a
stochastic process.

%
\begin{table}
\tablewidth=260pt
\caption{Comparison of the model fit to American put option data using
$\lambda= \lambda^*$ and $\lambda= 0$ in terms of the mean-squared
errors, $S(\lambda)$, between model-predicted option prices and market
observed option prices}\label{tab:lambdasse}
\begin{tabular*}{260pt}{@{\extracolsep{\fill}}lcc@{}}
\hline
\textbf{Equity} & $\bolds{S(\lambda= \lambda^*)}$ & $\bolds{S(\lambda= 0)}$ \\
\hline
Dell & 0.1839\phantom{0} & 0.7580\phantom{0} \\
Disney & 0.3631\phantom{0} & 0.9620\phantom{0} \\
Xerox & 0.02016 & 0.02739 \\
\hline
\end{tabular*}
\end{table}


\section{Discussion} 
\label{sec:discuss}

We introduce an algorithm for pricing American-style options under
stochastic volatility models, where volatility is assumed to be a
latent process. Our illustrative model takes into account the
co-dependence between share prices and their volatility as well as the
market price of volatility risk. The approach is based on (i) the
empirical observation that the conditional filtering distributions $\pi
_t$ can be well approximated by summary vectors $Q_t$ or parametric
families of distributions that capture their key features, (ii) the use
of a sequential Monte Carlo step to obtain and update the distributions
$\pi_t$, and (iii) a gridding (quadrature) type approach and a Monte
Carlo simulation-based adaptation to solve the associated dynamic
programming problem. Our methodology is not tied to a specific
stochastic volatility model or simulation procedure but could
accommodate a wide range of stochastic volatility models and/or
numerical simulation methods.

We document, through numerical experiments, that our method uses
features of $\pi_t$ to better price American options more accurately
than simpler methods. In fact, our approach comes within standard error
of the American option price when volatility is assumed to be observed.
One drawback with the methodology that we introduce is its
computational demand. Additionally, there are special situations (high
mean reversion and low volatility of volatility) where simpler methods
may suffice for pricing American-style options. However, our approach
leads to a more optimal exercise rule for all model/option parameter
settings. Our approach can also be practically implemented using
sophisticated parallel computing resources.
The proposed valuation method for pricing American-style options is
especially useful for important financial decisions in a very volatile
market period.

Using observed market data on share prices for three equities (Dell,
Disney and Xerox), we implement a Bayesian inferential procedure to
estimate (i) share price model parameters, and (ii) the market price of
volatility risk (or the volatility risk premium). Our results are
consistent with findings in the literature, namely, leverage effects
between share prices and their volatility and a negative volatility
risk premium. Leverage effects are significant for all equities with
the exception of Xerox.
The volatility risk premium (measured by $\lambda$) is also
significantly negative since its credible interval does not span zero
for any of the three equities.
This ultimately implies that volatility risk is priced in the market
and investors are willing to pay a premium for adverse movements in
share prices due to volatility. Furthermore, we approximate the
posterior distribution of $\lambda$ near its optimal value with a
Gaussian distribution. Consequently, we are able to make statistical
inference about the volatility risk premium for early-exercise options.

A potential refinement of our estimation procedure would be to
implement a joint time series analysis of the share and option prices.
This analysis can be facilitated by the algorithms in this paper,
however, parallel computing power would be of tremendous assistance in
this regard. Additionally, jumps in the statistical models could also
be incorporated and our approach could be used to make inference on
jump parameters and the jump risk premium. An additional line of future
work would be to use the inference made on the volatility risk premium
of American-style options to construct profitable trading/hedging
strategies that are pertinent to risk management settings.

\begin{appendix}

\section*{Appendix} 
\label{sec:appendix}
\subsection{\texorpdfstring{Proof of Lemma~\protect\ref{lem:1}}{Proof of Lemma 3.1}}
\label{subsec:lemmaproofs}
We use an inductive argument. To begin with,
$u_T(s_0,\ldots,s_T,d_T)$ is by its definition~(\ref{eq:uTdefn})
obviously a function of $s_T$ and $d_T$, and thus is trivially
a functional of $s_T,\pi_T$ and $d_T$, which we can denote
by $\tilde{u}_T(s_T,\pi_T,d_T)$.

Next, suppose that for some $t$, we can write
$u_{t+1}(s_0,\ldots,s_{t+1},d_{t+1}) = \break \tilde{u}_{t+1}(s_{t+1},\pi
_{t+1},d_{t+1}).$
Then from~(\ref{eq:dprecursion}),
%
\begin{equation}
\label{eq:lmeq1}
u_t(s_0,\ldots,s_t,E) = g(s_t),
\end{equation}
and
%
\begin{eqnarray}
&&u_t(s_0,\ldots,s_t,H)\nonumber\\
&&\qquad= E_{\mathrm{RN}}\bigl({u_{t+1}^*(s_0,\ldots,s_t,S_{t+1}) |
S_0=s_0,\ldots,S_t=s_t}\bigr) \nonumber
\\
\label{eq:lmeq2}
&&\qquad= E_{\mathrm{RN}}\bigl(\tilde{u}_{t+1}^*(S_{t+1},\pi_{t+1}) | S_0=s_0,\ldots,S_t=s_t\bigr)
\\
\label{eq:lmeq3}
&&\qquad= \int E_{\mathrm{RN}}\bigl(\tilde{u}_{t+1}^*(S_{t+1},\pi_{t+1}) | S_0=s_0,\ldots
,S_t=s_t,Y_t=y_t\bigr)\pi_t(y_t)\, dt
\\
\label{eq:lmeq4}
&&\qquad= \int E_{\mathrm{RN}}\bigl(\tilde{u}_{t+1}^*(S_{t+1},\pi_{t+1}) | S_t=s_t,Y_t=y_t\bigr)
\pi_t(y_t) \,dt.
\end{eqnarray}
Equation~(\ref{eq:lmeq3}) is obtained from~(\ref{eq:lmeq2}) using a simple
conditioning argument, and~(\ref{eq:lmeq4}) then follows since
$\{(S_t,Y_t),~t=0,1,\ldots\}$ is a (bivariate) Markov process.
The expression in~(\ref{eq:lmeq1}) is obviously a function of $s_t$,
and since $s_t$ and $\pi_t$ completely determine the distribution
of the arguments $S_{t+1}$ and $\pi_{t+1}$ to the function $\tilde
{u}_{t+1}^*(\cdot,\cdot)$
in~(\ref{eq:lmeq4}), it is also clear that the expression in~(\ref{eq:lmeq4})
is a functional of $s_t$ and $\pi_t$.
Thus, $u_t(s_0,\ldots,s_t,d_t)$ is a functional of $s_t$, $\pi_t$ and $d_t$,
which we denote by $\tilde{u}_t(s_t,\pi_t,d_t)$.

Invoking this inductive step for $t=T-1,T-2,\ldots,0$ gives the
first part of the desired result.
The second part of the result follows directly from the first part,
along with the definitions~(\ref{eq:utstardefn}) and~(\ref{eq:dtstardefn}).


\subsection{Estimation algorithms}\label{subsec:mcmcvolrisk}

The MCMC algorithm that we use to estimate the posterior distribution
$p(\theta|r_1, \ldots, r_n)$ in equation (\ref{eq:thetapost}) is
described below in Algorithm~\ref{alg:mcmc}. We implement a random-walk
Metropolis--Hastings (MH) algorithm to arrive at our estimate of the
posterior distribution of $\theta$. Additionally, Algorithm~\ref
{alg:volrisk} describes the procedure used to optimize the
(approximate) posterior distribution of the volatility risk premium
$p(\lambda|u_1, \ldots, u_L)$ in equation (\ref{eq:lambdapost}).

%
\begin{algorithm}[t]
\caption{Markov chain Monte Carlo (MCMC) posterior simulation}
\label{alg:mcmc}
\begin{algorithmic}[0]
\STATE\textit{Initialization 1.} Input the parameters of the prior and\vspace*{1pt}
proposal distributions corresponding to the $\tilde{\theta}$ parameterization.
\vskip2mm
\STATE\textit{Initialization 2.} Set the starting value of $\tilde
{\theta
}$ at the prior mean (or any other reasonable value) and denote this by
$\tilde{\theta}_c$ to represent the current value. Use Algorithm~\ref
{alg:kitagawa} and the log-prior densities to compute a log-posterior
value of $\tilde{\theta}_c$ and denote this by $\mbox{LPost}_c$.
\vskip2mm
\For{$i=1,\ldots,B$}
\begin{itemize}[\quad]
\item\textit{Sampling.} Draw the $i$th potential value of $\tilde{\theta}$
using a multi-variate normal proposal density and denote this by
$\tilde
{\theta}_s$.
\item\textit{Posterior evaluation.} Use Algorithm~\ref{alg:kitagawa} along with
the log-prior densities to compute the $i$th log-posterior value of
$\tilde{\theta}_s$ and denote this by $\mbox{LPost}_s$.
\item\textit{MH-step A.} Sample $U_i \sim\operatorname{Unif}[0,1]$.
\item\textit{MH-step B.} If $\log(U_i) \leq(\mbox{LPost}_s - \mbox{LPost}_c)$,
update $\tilde{\theta}_c = \tilde{\theta}_s$ and $\mbox{LPost}_c =
\mbox{LPost}_s$. Else do not update $\tilde{\theta}_c$ and $\mbox{LPost}_c$.
\end{itemize}
\EndFor
\vskip2mm
\textit{Output.} Perform the relevant inverse transformation of
$\tilde{\theta}$ in order to return the posterior distribution of
$\theta$.
\end{algorithmic}
\end{algorithm}

\begin{remark*}
We implement in Algorithm~\ref{alg:mcmc} a stochastic search over the
transformed parameter space $\tilde{\theta}$ as defined in
Section~\ref
{subsec:modelparameters}.
We use a multivariate Gaussian proposal density with mean equal to the
current point and a diagonal variance--covariance (VCOV) matrix. The
elements of the VCOV matrix that proposed values for $\rho$, $\alpha$,
$\beta$ and $\gamma$ are, respectively,
\[
0.001,\qquad 0.005,\qquad 0.0025\quad \mbox{and}\quad 0.001.
\]
(We experimented with different parameterizations of the proposal
density and did not find appreciable differences in the results.)
\end{remark*}

\begin{remark*}
Regarding the statistical estimation of the model parameters via MCMC,
we initialize the physical drift rate using the average of the returns
data and we set the correlation parameter $\rho=0$ for each case. We
initialize the parameters of the stochastic volatility process $(\alpha
, \beta, \gamma)$ as follows:
Dell $(8.20, -1.00, 1.50)$, Disney $(4.40, -1.20, 1.10)$ and Xerox
$(17.0, -0.800, 3.00)$.
These are approximate maximum likelihood estimates that are computed
using Cronos, an open source software written by Anthony Brockwell and
available at \url{http://www.codeplex.com/cronos}.
\end{remark*}

\begin{remark*}
We set $B$ to 50,000 and take a burn-in period of 5000 in
Algorithm~\ref{alg:mcmc}. Convergence is ascertained using trace plots
of the posterior output.
\end{remark*}

\begin{remark*}
We use the Monte Carlo based approach described in Section~\ref
{sec:pricealg} to evaluate $P_{\theta^*}^i(\lambda_j)$, as we found
this to be faster for the purposes of our empirical analysis in
Algorithm~\ref{alg:volrisk}.
\end{remark*}


\subsection{Calculation of posterior distribution of $\lambda$}
\label{subsec:lambdapost}

We now outline some of the calculations that are needed to compute the
normal approximation [equation (\ref{eq:lambdanormal})] to the
posterior distribution of $\lambda$ near the mode of its true posterior
distribution [equation (\ref{eq:lambdapost})]. Recall that the mean of
the normal approximation is the posterior mode. The reciprocal of the
variance term is
\[
\frac{1}{V^*} = -\frac{d^2 \log (p(\lambda|u_1, \ldots, u_L)
)}{d\lambda^2} \bigg|_{\lambda= \lambda^*} = -\frac{d^2 \log
[(S(\lambda))^{-L/2} ]}{d\lambda^2} \bigg|_{\lambda= \lambda^*}.
\]
Observe that
\[
-\frac{d^2 \log [(S(\lambda))^{-L/2} ]}{d\lambda^2} = \frac{L}{2}
\biggl[\frac{S''(\lambda) \cdot S(\lambda) - (S'(\lambda))^2}{(S(\lambda
))^2} \biggr].
\]
We approximate the first and second derivative expressions $S'(\lambda
)$ and $S''(\lambda)$ as\footnote{Theoretically, the first derivative
is equal to 0 at the optimized point [i.e., $S'(\lambda^*)=0$]. The
numerical approximation of the first derivative using the expressions
above comes within tolerance of~0.}
\begin{eqnarray*}
S'(\lambda) &\approx&\frac{S(\lambda+ \Delta_G) - S(\lambda- \Delta
_G)}{2 \Delta_G}\quad \mbox{and}
\\
S''(\lambda) &\approx&\frac{S(\lambda+ \Delta_G) -2S(\lambda) +
S(\lambda- \Delta_G)}{\Delta_G^2}.
\end{eqnarray*}
In order to evaluate the derivative expression at $\lambda^*$, we use
the values in Table~\ref{tab:lambdapostcalc}. Once these computations
are completed, the normal approximation to the posterior distribution
of $\lambda$ is completely specified.

\begin{table}
\tablewidth=255pt
\caption{Values of $S(\lambda)$ at different points. Recall the values
for $\lambda^*$ are given in Table~\protect\ref{tab:lambdaresults}.
Additionally, the spacing between the $\lambda$-grid, $\Delta_G$,
equals 0.05}\label{tab:lambdapostcalc}
\begin{tabular*}{255pt}{@{\extracolsep{\fill}}lccc@{}}
\hline
\textbf{Equity} & $\bolds{S(\lambda^* - \Delta_G)}$ & $\bolds{S(\lambda^*)}$ & $\bolds{S(\lambda^* + \Delta_G)}$ \\
\hline
Dell & 0.1847\phantom{0} & 0.1839\phantom{0} & 0.1860\phantom{0} \\
Disney & 0.3658\phantom{0} & 0.3631\phantom{0} & 0.3644\phantom{0} \\
Xerox & 0.02023 & 0.02016 & 0.02030 \\
\hline
\end{tabular*}
\end{table}


\subsection{Numerical simulation results}
\label{subsec:euler}

%
\begin{algorithm}
\caption{Posterior analysis of the market price of volatility risk}
\label{alg:volrisk}
\begin{algorithmic}[0]
\STATE\textit{Initialization 1.} Compute the posterior summary from the
model estimation routine outlined in Algorithm~\ref{alg:mcmc}. (This
could be, for example, the posterior mean or median of $\theta$.)
Denote the posterior summary measure by $\theta^*$.
\vskip2mm
\STATE\textit{Initialization 2.} Input the $L$ American option contract
features including initial share price and initial volatility based on,
say, a 10-day historical volatility measure.
\vskip2mm
\STATE\textit{Initialization 3.} Input a grid of $\lambda$ values that
will be used to find the optimal value for $\lambda$. (This can be
roughly estimated via trial and error.) Denote the number of grid
points by $G$ and the $\lambda$ values by $\lambda_1, \ldots,
\lambda
_G$ and the distance between each grid point by $\Delta_G$.
\vskip2mm
\For{$i=1, \ldots, G$}
\begin{itemize}[\quad]
\item\textit{Option valuations.} For $j=1,\ldots,L$, compute and store the
model-predicted American option values, $P_{\theta^*}^j(\lambda_i)$
using the pricing algorithms (either Monte Carlo or grid-based
described in Section~\ref{sec:pricealg}).
\item\textit{Optimize SSE.} Compute the value of the sum of squared errors
$S(\lambda)$ defined in equation (\ref{eq:sse}).
\end{itemize}
\EndFor
\vskip2mm
\textit{Find optimal $\lambda$.} Find the optimal $\lambda$ value,
$\lambda^*$, among the grid points $(\lambda_1, \ldots, \lambda_G)$
such that
\[
\lambda^* = \mathop{\arg\min}_{\lambda} S(\lambda).
\]
\vskip2mm
\textit{Posterior computation.} Starting with the prior
specification in equation (\ref{eq:priors}),
$S(\lambda)^{-L/2}$ is the posterior distribution of $\lambda$
up to a constant of proportionality. Calculate the approximate
posterior distribution by using the Gaussian approximation to the
posterior distribution near the mode [i.e., near $\lambda^*$; see \citet{sebwild03} or
\citet{Schervish95}].
\vskip2mm
\textit{Output.} Return the approximate posterior distribution of
$\lambda$ from equation (\ref{eq:lambdanormal}) and summarize
accordingly. (Derivative evaluations are evaluated numerically using
central difference methods.)
\end{algorithmic}
\end{algorithm}

Tables~\ref{tab:lsmeuler} and~\ref{tab:smcfulleuler} provide the
results of the numerical experiments, described in Section~\ref
{subsec:numerical}, when using an Euler discretization from the
stochastic volatility model in equations (\ref{eq:rnsvmodelall1}),
(\ref{eq:rnsvmodelall2}) and (\ref{eq:rnsvmodelall3}).
As can be observed from the results in these tables, the option prices
using the Euler discretization are almost identical to those using the
exact simulation for all methods described in Section~\ref
{subsec:numerical}. Hence, when working with models that may not permit
exact simulation, first and higher order discretization techniques also
facilitate option pricing methods such as the ones done in this analysis.

%
\begin{table}[b]
\caption{Euler discretization---American put option pricing results
(and compute times) using methods \textup{A} (basic LSM with past share prices)
and \textup{B} (realized volatility as an estimate for volatility). Results are
very similar to those reported for the exact simulation in Table~\protect\ref{tab:lsm}}\label{tab:lsmeuler}
\begin{tabular*}{\textwidth}{@{\extracolsep{\fill}}lcc@{}}
\hline
\textbf{Experiment no.} & \textbf{A (basic LSM)} & \textbf{B (realized volatility)} \\
\hline
1 & 3.047 (0.0107) & 3.040 (0.0103) \\
Time (sec) & 8 & 8 \\
2 & 2.147 (0.00946) & 2.154 (0.00974) \\
Time (sec) & 15 & 15 \\
3 & 1.213 (0.00759) & 1.230 (0.00841) \\
Time (sec) & 13 & 11 \\
4 & 3.569 (0.0242) & 4.152 (0.0358) \\
Time (sec) & 37 & 39 \\
5 & 12.995 (0.0743) & 15.066 (0.120) \\
Time (sec) & 36 & 42 \\
6 & 18.623 (0.121) & 21.474 (0.168) \\
Time (sec) & 44 & 44 \\
7 & 1.588 (0.0120) & 1.843 (0.0186) \\
Time (sec) & 13 & 14 \\
8 & 0.0945 (0.00367) & 0.145 (0.00553) \\
Time (sec) & 16 & 14 \\
9 & 2.437 (0.0132) & 2.668 (0.0197) \\
Time (sec) & 20 & 21 \\
\hline
\end{tabular*}
\end{table}

%
\begin{table}
\caption{Euler discretization---American put option pricing results
(and compute times) using methods \textup{C} (MC/Grid) and \textup{D} (observable
volatility). Results are very similar to those reported for the exact
simulation in Table~\protect\ref{tab:smcfull}}\label{tab:smcfulleuler}
\begin{tabular*}{\textwidth}{@{\extracolsep{\fill}}lcc@{}}
\hline
\textbf{Experiment no.} & \textbf{C (MC/Grid)} & \textbf{D (observable volatility)} \\
\hline
1 & 3.052 (0.0109) & 3.045 (0.00982) \\
Time (sec) & 161 & 8 \\
2 & 2.164 (0.00958) & 2.173 (0.00986) \\
Time (sec) & 321 & 15 \\
3 & 1.257 (0.00900) & 1.261 (0.00902) \\
Time (sec) & 212 & 10 \\
4 & 4.684 (0.0439) & 4.734 (0.0441) \\
Time (sec) & 761 & 37 \\
5 & 16.270 (0.138) & 16.330 (0.138) \\
Time (sec) & 748 & 37 \\
6 & 22.764 (0.183) & 22.999 (0.185) \\
Time (sec) & 831 & 47 \\
7 & 1.998 (0.0217) & 2.045 (0.0225) \\
Time (sec) & 302 & 16 \\
8 & 0.169 (0.00691) & 0.172 (0.00704) \\
Time (sec) & 236 & 15 \\
9 & 2.851 (0.0233) & 2.861 (0.0235) \\
Time (sec) & 379 & 24 \\
\hline
\end{tabular*}
\end{table}

\end{appendix}

\section*{Acknowledgments}

The authors are grateful to John Lehoczky, Mark Scher\-vish, Duane Seppi,
Nikunj Kapadia, Daniel Peris and Michael Sullivan for a number of
useful conversations related to the work in this paper. They are also
grateful to the Editor, Associate Editor and anonymous referees whose
suggestions have markedly improved the content of this work.
B. R. Rambharat thanks Robert
Wolpert and Merlise Clyde for granting access to the Duke Shared
Cluster Resource at the Center for Computational Science, Engineering
and Medicine to work on a portion of this research. Finally, we are
very thankful for the patience and assistance of Pantelis Vlachos and
the Remarks Computing Group in the Department of Statistics at Carnegie
Mellon University.


%
\begin{supplement}[id=suppA]
\stitle{Code and data sets}
\slink[doi]{10.1214/09-AOAS286SUPP}
\slink[url]{http://lib.stat.cmu.edu/aoas/286/supplement.zip}
\sdatatype{.zip}
\sdescription{\textit{Sequential Monte Carlo pricing
routines}. The R code used in our analysis for pricing American-style
options in a latent stochastic volatility framework as well code for
optimizing all model parameters, including the market price of
volatility risk, are part of this supplement. \textit{American put
option data sets}. The data sets used in our pricing/estimation
analysis include historical share prices and American put option prices
for three equities: Dell, Disney and Xerox. The data files are in this
supplement.}
\end{supplement}

\printaddresses

\end{document}